\documentclass[conference]{IEEEtran}
\IEEEoverridecommandlockouts
\usepackage{cite}
\usepackage{amsmath,amssymb,amsfonts,amsthm}
\usepackage{algorithmic}
\usepackage{graphicx}
\usepackage{textcomp}
\usepackage{xcolor}
\def\BibTeX{{\rm B\kern-.05em{\sc i\kern-.025em b}\kern-.08em
    T\kern-.1667em\lower.7ex\hbox{E}\kern-.125emX}}

\usepackage{amsmath,amssymb,epsfig,xspace,xcolor,subfig}

\usepackage{accents}
\usepackage{enumitem}
\usepackage{url}

\usepackage[T1]{fontenc}

\usepackage{todonotes}

\usepackage[lined,ruled,vlined,noend]{algorithm2e}
\DontPrintSemicolon 
\LinesNumbered

\newcommand{\lb}{\linebreak[0]}

\xspace
\xspace
\xspace
\xspace
\xspace
\xspace
\xspace
\xspace
\newcommand\ci[2]{\ensuremath{#1[#2]}}\xspace
\newcommand\spt{\ensuremath{\text{\emph{sub\_cell}}}}\xspace
\newcommand\pdc{\ensuremath{\Gamma}}\xspace
\newcommand\pdom{\ensuremath{\preceq}}\xspace
\newcommand\dom{\ensuremath{\prec}}\xspace
\newcommand\sdom{\ensuremath{\precsim}}\xspace
\xspace

\newcommand\cps{\ensuremath{\mathcal{CA}}\xspace}
\newcommand\cs{\ensuremath{\mathcal{C}}\xspace}
\newcommand\gs{\ensuremath{\mathcal{G}}\xspace}

\newcommand\ks{\ensuremath{\mathcal{K}}\xspace}
\newcommand\rs{\ensuremath{\mathcal{R}}\xspace}
\newcommand\ls{\ensuremath{\mathcal{L}}\xspace}

\newcommand\ps{\ensuremath{\mathcal{P}}\xspace}
\newcommand\ts{\ensuremath{\mathcal{L}}\xspace}

\newcommand\sss{\ensuremath{\mathcal{S}}\xspace}

\newcommand\bbii{\ensuremath{\mathcal{I}}\xspace} 
\newcommand\ds{\ensuremath{\bold{\beta}}\xspace} 
\newcommand\bo{\ensuremath{\mathcal{O}}\xspace} 

\newcommand\lcw[1]{{\color{blue} #1}}
\newcommand\tu[1]{\ensuremath{\mathcal{T}[#1].u}}
\newcommand\tl[1]{\ensuremath{\mathcal{T}[#1].l}}
\newcommand\too[1]{\ensuremath{\mathcal{T}[#1].o}}
\newcommand\tul[1]{\ensuremath{\mathcal{T}[#1]}}
\newcommand\ttt{\ensuremath{\mathcal{T}}}

\newtheorem{theorem}{Theorem}
\newtheorem{corollary}{Corollary}
\newtheorem{lemma}{Lemma}
\newtheorem{definition}{Definition}

\begin{document}

\title{SkyCell: A Space-Pruning Based Parallel Skyline Algorithm}
\author{%
  {Chuanwen Li{\small $~^{\#}$}, Yu Gu{\small $~^{\#}$}, Jianzhong Qi{\small
      $~^{*}$}, Ge Yu{\small $~^{\#}$}}%
  \vspace{1.6mm}\\
  \fontsize{10}{10}\selectfont\itshape
  $^{\#}$\,College of Computer Science and Engineering, Northeastern
  University, China\\
  \fontsize{9}{9}\selectfont\ttfamily\upshape
  {\{lichuanwen, guyu, yuge\}@mail.neu.edu.cn}
  \vspace{1.2mm}\\
  \fontsize{10}{10}\selectfont\rmfamily\itshape
  $^{*}$\,School of
  Computing and Information Systems, The University of Melbourne, Australia\\
  \fontsize{9}{9}\selectfont\ttfamily\upshape
  jianzhong.qi@unimelb.edu.au
}

\maketitle

\begin{abstract}
  Skyline computation is an essential database operation that has many applications in multi-criteria decision making scenarios such as recommender systems. Existing algorithms have focused on checking point domination, which lack  efficiency over large datasets. We propose a grid-based structure that enables grid cell domination checks. We show that only a small constant number of cells need to be checked which is independent from the number of data points. Our structure also enables parallel processing. We thus obtain a highly efficient parallel skyline algorithm named \emph{SkyCell}, taking advantage of the parallelization power of graphics processing units. Experimental results confirm the effectiveness and efficiency of SkyCell -- it outperforms state-of-the-art algorithms consistently and by up to over two orders of magnitude in the computation time.
\end{abstract}


\section{Introduction}
\label{sec:intro}

The \emph{skyline query} is an essential query in multi-criteria decision making applications such as recommender systems and  business management (to compute the \emph{Pareto frontier})~\cite{bogh2015work,borzsony2001skyline,chomicki2003skyline,kohler2011efficient,liu2018skyline, yu2019efficient}. It retrieves data points that are not \emph{dominated} by any other points in a dataset. 
Suppose each point has $d$ attributes. A point $p_i$  dominates another point $p_j$ if $p_i$ is better than $p_j$ in at least one attribute and is as good as $p_j$ in all other attributes. The ``better than'' relationship is often quantified as having a smaller attribute value. 
Consider recommending restaurants to a user. In Table~\ref{tab:epl}, there are four restaurants each with three attributes: average cost per person, distance to the user, and rating rank (1 is the top rank). Restaurants $r_1$ and $r_3$ are dominated by $r_2$, as they are more expensive, farther away, and rated lower.  Neither $r_2$ nor $r_4$ is dominated. They are the skyline points, which can be used 
for recommendation.

\begin{table}[h]
  \centering
  \caption{A Restaurant Recommendation Example}
  \label{tab:epl}
  \begin{tabular}{c|c c c}
    \hline
    Restaurant&Average Cost &Distance&Rating Rank\\
    \hline
    \hline
    $r_1$&\$12&9 km&$3$\\
    $r_2$&\$8&3 km&$2$\\
    $r_3$&\$10&17 km&$4$\\
    $r_4$&\$26&8 km&$1$\\
    \hline
  \end{tabular}
\end{table}

Existing skyline algorithms mostly fall into two groups: \emph{sorting-based} and \emph{partitioning-based}\cite{zou2008novel}. 
Both groups maintain a \emph{skyline buffer} that stores the skyline points. They grow the buffer by comparing points outside the buffer with those inside. Sorting-based algorithms rearrange the dataset such that skyline points are more likely to be processed and added to the buffer early on. This helps the point elimination efficiency. 
Partitioning-based algorithms structure the skyline points in the buffer such that the remaining points each only needs to compare against a subset of the skyline points.


On-going efforts~\cite{bogh2015work, wang2019scalable, bogh2016skyalign, islam2017computing, zois2019complex, lougmiri2017new} have been made to parallelize skyline computation. \emph{Graphics processing units} (GPU) are used for their strong parallelization capability. Most GPU-powered sorting-based skyline algorithms~\cite{choi2012multi, bogh2013efficient} are adaptations of their sequential counterparts. 
These algorithms also check all points to grow the skyline buffer, which hinders their efficiency. The state-of-the-art sorting-based algorithm~\cite{liu2018skyline} turns back to sequential processing. 
 This algorithm, however, requires expensive pre-computations and may suffer when there are  updates. Partitioning-based algorithms have recursive partitioning procedures~\cite{zhang2009scalable, nasridinov2017two, lee2014scalable, lee2017efficient} or tree-like structures to reduce point domination checks~\cite{bogh2015work}. 
 They are intrinsically difficult for GPU processing. To avoid such issues, the state-of-the-art partitioning-based skyline algorithm uses GPU with a grid partition~\cite{bogh2015work}. It partitions each dimension into 16 segments regardless of the dataset size, which cannot fully exploit the GPU throughput and may cause branch divergence of GPU warps.

 A key limitation in the existing algorithms is that they mostly check for \emph{point} domination (or \emph{point-partition} domination, detailed in Section~\ref{sec:rlwk}) to identify the skyline points. They lack efficiency as the number of data points becomes large. For example, OpenStreetMap has billions of  points~\cite{osm_stat}. Computing the skyline points from data in such a scale takes some 10 seconds even with the state-of-the-art GPU-based parallel algorithm~\cite{bogh2015work} (detailed in Section~\ref{sec:ee}). This hinders user experience for online skyline queries (e.g., over dynamic data with updates). We aim to achieve sub-second skyline query time on such data.

We observe that the data space can be partitioned into regions such that domination checks   can be performed among the regions. This enables pruning by regions without examining the  points in each region. We show that only a small constant number of (non-dominated)  regions contain skyline points. We thus propose an efficient algorithm to compute such regions and hence the skyline points, which scales much better with the dataset size. 

We partition the data space with a regular grid and check for domination between 
the grid cells based on their relative positions. Intuitively, the cells with smaller coordinates dominate those with larger ones. We show that only cells that are \emph{not} dominated 
 contain skyline points. Such cells are named the \emph{candidate cells}. 

We prove that the number of candidate cells is bounded by the data dimensionality 
and the grid granularity, and it is \emph{independent} of the dataset size. 
We further show that a candidate cell can be partitioned recursively to form smaller candidate cells (in grids of larger granularities). As the grid granularity becomes larger, 
each candidate cell becomes smaller, and the potion of the space covered by candidate cells decreases monotonically.  For an $8\times 8$ grid in two-dimensional Euclidean space, there are only 15 candidate cells (i.e., 23\% of the 64 cells). When the granularity increases to $32 \times 32$, there are only 63 candidate cells (i.e., 6.2\% of the 1,024 cells). 

Based on these key properties, we proposed a \textbf{cell}-based \textbf{sky}line algorithm named \textbf{SkyCell} that progressively computes the candidate cells in grids with increasing granularities, until each candidate cell contains only a small number of points. From the resultant cells, skyline points can be computed efficiently with existing point domination based algorithms (e.g., sort-first skyline~\cite{chomicki2003skyline}). 

 SkyCell processes each candidate cell independently. This offers an important opportunity to improve the algorithm efficiency with parallelization. We thus further propose a parallel SkyCell
algorithm using GPU. 
To take full advantage of the parallelization power of GPU, we carefully design our algorithm to avoid warp divergence, and we arrange the data to promote coalesced memory access. 
We thus achieve a highly efficient algorithm that outperforms state-of-the-art parallel skyline algorithms by up to two orders of magnitude. 

In summary, we make the following contributions: 
\begin{itemize}
\item We propose a novel approach for skyline computation based on grid partitioning and candidate cells. By using cell domination checks, our approach significantly reduces the number of domination checks, thus yielding a much better scalability to the dataset size.
\item We derive a theoretical bound on the number of candidate cells to be examined. We further show how such cells can be recursively partitioned to yield smaller cells without missing any skyline points. Based on these, we propose a skyline algorithm named SkyCell.

\item Since the candidate cells can be computed independently,  we further propose 
a parallel SkyCell algorithm, taking full advantage of the parallelization power of GPU. Note that our algorithms do not require any pre-computation. Thus, they are also robust to data updates. 

  
\item We perform cost analysis and extensive experiments. The results confirm the superiority of our algorithm over the state-of-the-art parallel and sequential skyline algorithms.
\end{itemize}


\section{Related Work}
\label{sec:rlwk}

The skyline query was first studied in computational geometry and was called the  \emph{maxima}~\cite{kung1975finding}. It was later introduced to the database community and was extensively studied~\cite{borzsony2001skyline,hose2012survey, tan2001efficient, kossmann2002shooting, huang2006skyline, kulkarni2019skyline}. 
Below, we review the representative sequential and parallel algorithms.

\subsection{Sequential Skyline Algorithms}
\label{sec:st}

The \emph{block-nested-loops} (BNL)~\cite{borzsony2001skyline} algorithm forms the basis of skyline computation. It processes the points sequentially and keeps track of the points that are not dominated by any other points seen so far in a \emph{skyline buffer} $C$. When a  point $p$ is processed, it is compared against the points in $C$. If $p$ is dominated by some point in $C$, it is skipped.  Otherwise, $p$ is added to $C$, and existing points in $C$ that are dominated by $p$ is removed from $C$.

The \emph{sort-first skyline} (SFS)~\cite{chomicki2003skyline} algorithm optimizes BNL by sorting the points first (by Manhattan norm). By the sorted order, once a point is added to the skyline buffer, it will not be dominated by points added later. 
Another study~\cite{lee2007approaching} uses the Z-order for sorting.
The \emph{branch-and-band skyline} (BBS)~\cite{papadias2005progressive} algorithm constructs an R-tree and pre-computes the \emph{mindist} of intermediate entries for skyline  pruning.  When a tree node is visited, only child nods on its lower-left may contain skyline points and need to be visited. 
{These two works~\cite{lee2007approaching,papadias2005progressive} also prune by partitions, but they use point-partition domination checks. 
Lee et al.~\cite{lee2007approaching} use points on a Z-curve to prune partitions. The ordered pruning process makes it difficult to parallelize. BBS~\cite{papadias2005progressive} prunes a partition $c$ by checking whether there are points in another partition $c'$, which is a partition inside which any point dominates all points in $c$. BBS also needs to visit the points orderly and hence is difficult to parallelize.}

Another series of studies takes a space partitioning approach. \emph{Voronoi-based spatial skyline} (VSS)~\cite{sharifzadeh2006spatial} builds a Voronoi diagram over the data space to answer \emph{spatial skyline queries} (SSQ). 
SSQ aims to return skyline points based on attributes constructed online.  
In an SSQ, 
there are a set of $d$ query points, and the $d$ 
attributes of a data point $p$ are computed online 
as the distances between $p$ and the query points.  
 VSS  visits the points in a best-first order based on their distances to the query points, starting from a point closest to any one of the query points. 
When a point $p$ is visited, its Voronoi neighbors that pass a validity test 
are added to the list of points to be visited next. 
Further, if $p$ is not dominated by any skyline points found so far, it is added to the skyline set. 
  \emph{Skyline diagram} (SD)~\cite{liu2018skyline} pre-computes a Voronoi-like diagram.  
Query points falling in the same cell in the diagram will have the same skyline points, which are pre-computed.  When processing a skyline query, SD only needs to locate the cell that encloses the query point to fetch the query answer. This algorithm may suffer in pre-computation and storage costs when there are many skyline points. 

\subsection{Parallel Skyline Algorithms}
\label{sec:dp}


There are also many parallel skyline algorithms~\cite{balke2004efficient, zois2018massively, kohler2011efficient, zhu2017parallelization, bogh2017template, wang2019scalable}. The \emph{GPU-based Nested Loop} (GNL)~\cite{choi2012multi} algorithm is a parallel extension of BNL. 
It assigns a thread for each point and checks the point with all other points in parallel. 
 \emph{GPGPU Skyline} (GGS)~\cite{bogh2013efficient} sorts the points by the Manhattan norm. It then runs domination checks in multiple iterations. In each iteration, GGS uses the top-ranked unchecked points as the skyline buffer and compare them against the other points  in parallel. The non-dominated points in the skyline buffer are added to the skyline set. The dominated points and those added to the skyline set are excluded from future iterations. The process repeats until all points are processed. 

The \emph{balanced pivot selection} (BPS)~\cite{lee2014scalable, zhang2009scalable} algorithm uses GPU for pivot selection. It selects a pivot -- the point with the smallest normalized attribute values -- to split the data space into \emph{incomparable regions}. 
Points in different incomparable regions do not dominate each other.  
Each region is further split recursively.  
Pivots in the lower-level incomparable regions are computed  in parallel. 
Points are assigned to regions by comparing against the 
pivots, and they are only checked for domination in their assigned regions.

\emph{SkyAlign}~\cite{bogh2015work} is a GPU-based algorithm that uses a global, static partitioning scheme. It uses controlled branching to exploit transitive relationships between points and can avoid some  point domination checks. 
 It does \emph{not} use region-based domination checks, and it 
 has a fixed number of partitions regardless of the dataset size, which cannot make full use of the GPU throughput and may cause branch divergence of GPU warps.

A few other studies use MapReduce~\cite{mullesgaard2014efficient, park2013parallel, zhang2015efficient}. They focus on workload balancing among the worker machines. 


The main difference between the studies above and ours is that they focus on point domination checks, while we partition the space and check domination between the partitions, thus yielding significantly fewer domination checks and higher efficiency. 


\section{Preliminaries}
\label{sec:prel}

Given a set $\ps = \{p_1, p_2, \ldots, p_n\}$ of $n$ points in $d$-dimensional ($d > 1$) Euclidean space, 
we aim to compute the subset 
$\sss \subset \ps$ of all \emph{skyline points} in $\ps$, i.e., 
the \emph{skyline set} of $\ps$. 
Below, we define skyline points and key concepts. We list 
frequently used symbols in Table~\ref{tab:symbols}.


Skyline points are defined based on \emph{point domination}. Let $p[k]$ be the coordinate of a point $p$ in dimension $k$. 

\begin{definition}(Point domination)
  We say that a point $p_i$ \emph{dominates} another point $p_j$, denoted by $p_i \dom p_j$, if $\forall k \in [0, d), p_i[k] \le p_j[k]$ and $\exists l \in [0, d), p_i[l] < p_j[l]$. 
\end{definition}
\begin{definition}(Skyline point)
  We call $p_i \in \ps$ a \emph{skyline point} of $\ps$ if $p_i$ is not dominated by any other point $p_j \in \ps$, i.e., $\nexists p_j \in \ps, p_j \dom p_i$. 
\end{definition}

\begin{table}
  \caption{Frequently Used Symbols}
  \centering \footnotesize \renewcommand{\arraystretch}{1.2} {
    \begin{tabular}{p{0.5in}p{2.5in}}
      \hline
      Notation & Description \\
      \hline
      \hline
      $\ps$ & Data point set\\ 
      $\sss$ & Skyline set\\
      $d$ & Data dimensionality\\ 
      $p_i$ & A data point\\
      $p_i[k]$ & The coordinate of point $p_i$ in dimension $k$\\
      $p_i \dom p_j$ & Point $p_i$ dominates point $p_j$\\ 
      $\rho$ & The number of layers in our grid structure\\
      $\ls_i$ & The set of cells in Layer $i$\\
      $c$ & A cell in the grid structure\\
      $cl_k$ (or $c[k]$) & The dimension-$k$ index (column number) of a cell c\\
      $\cs_i$ & The set of candidate cells in Layer $i$\\ 
      $\ks_i$ & The set of key cells in Layer $i$\\
      $\lambda_i$ & An auxiliary point\\
      $\Lambda_i$ & The auxiliary key cell corresponding to $\lambda_i$\\
      $\spt(C)$ & The set of cells in the next layer from splitting the cells in $C$\\
      \hline 
    \end{tabular}
  }
  \label{tab:symbols}
\end{table}

Existing studies mainly focus on point (or point-partition) domination. 
We check for domination between space partitions. 
If a partition is  dominated, all points inside can be pruned.   
Next, we describe our structure to enable this partition-based pruning.

\begin{figure}[h]
  \centering
  \includegraphics[width=3.3in]{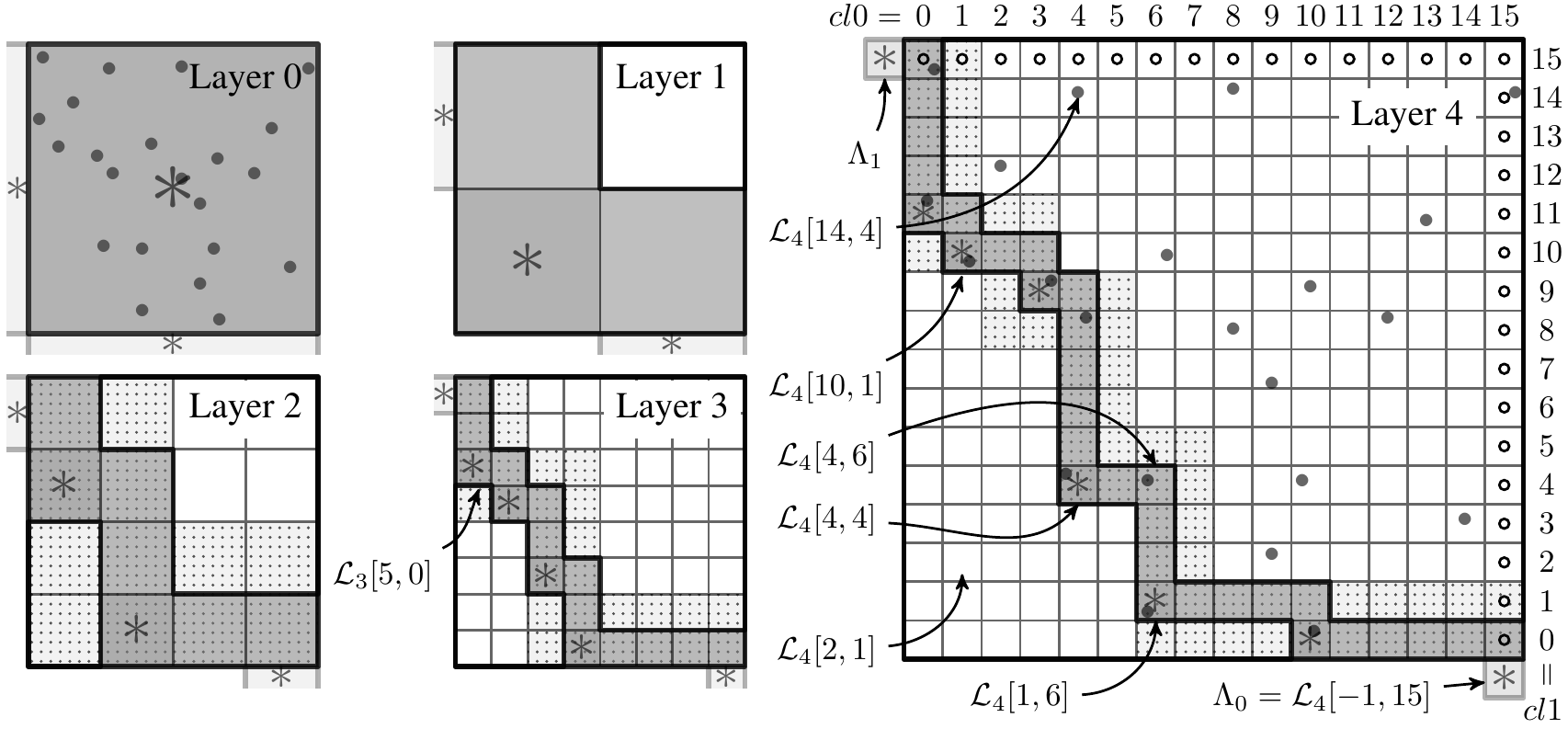}
  \caption{Example of multi-layered data space partitioning (the gray points denote data points)}
  \label{fig:egcells}
\end{figure}

\textbf{Our grid structure.} We consider the space as a $d$-dimensional unit hyper-cube and partition it with a multi-layer grid. The top grid layer (Layer~0) has the coarsest  granularity (i.e., the entire data space is a cell),  
while the bottom layer (Layer $\rho$, where $\rho$ is a system parameter) 
has the finest granularity. Each layer is a regular grid, 
with $2^{i \cdot d}$ \emph{cells} in Layer~$i$. 
In Fig.~\ref{fig:egcells}, $d = 2$, and we have 
$2^{0\times 2} = 1$ to $2^{4\times 2} = 256$ cells for Layers 0 to 4. 
Each layer has the same unit size. Layer 4 has been zoomed in 
for better visibility.  

Let the set of cells in Layer $i$ be $\ls_i$. A cell $c = \ls_i[cl_{d-1}, \ldots,  cl_0]$ is indexed by its column numbers, i.e., it is at columns $cl_{d-1}, \ldots,  cl_0$ in dimensions $d-1, \ldots,  0$, respectively. We use $c[k]$ to denote the index (column number) of $c$ in dimension $k$: $c[k] = cl_k$. In Fig.~\ref{fig:egcells}, 
 cell $c = \ls_4[10,1]$ in Layer 4 is at column 10 in dimension 1 (the vertical dimension) and column 1 in dimension 0 (the horizontal dimension), i.e., $c[1]= 10$ and $c[0]= 1$.

Since we consider points in a unit hyper-cube $[0, 1)^d$, in Layer $i$, the cell $c$ to which a point $p$ belongs is  calculated by:
\begin{equation}\label{eq:cell_column_number}
  c = \ls_i[\lfloor p[d-1]\cdot 2^{i} \rfloor, \ldots , \lfloor p[0]\cdot 2^{i} \rfloor]
\end{equation}
For example, in Fig.~\ref{fig:egcells}, point $p=(0.63, 0.08)$  belongs to cell 
$\ls_3[\lfloor 0.63\times 2^{3} \rfloor,\lfloor 0.08 \times 2^{3} \rfloor] = \ls_3[5, 0]$ 
in Layer 3 and cell $\ls_4[\lfloor 0.63\times 2^{4} \rfloor,\lfloor 0.08 \times 2^{4} \rfloor] = \ls_4[10, 1]$ in Layer 4. 

\textbf{Cell domination.} We prune based on   
cell domination in each layer. 
\emph{In what follows, when multiple cells are discussed, they refer to cells from the same layer, unless  otherwise stated.} 

\begin{definition}\label{df:cd}(Cell domination)
  We say that cell $c_i$ \emph{dominates} cell $c_j$, denoted by $c_i \dom c_j$, if $c_i$ is not empty (i.e.,  enclosing points in $\mathcal{P}$), and the index of $c_i$ is less than that of $c_j$ in each dimension, i.e.,
    \begin{equation}
  c\neq \emptyset \wedge \forall k \in [0, d), \ci{c_i}{k} < \ci{c_j}{k}
  \end{equation}
  We say that $c_i$ \emph{partially dominates} $c_j$, denoted by $c_i \pdom c_j$, if $c_i$ is not empty, the index of $c_i$ equals to that of $c_j$ in at least one dimension, 
  and the index of $c_i$ is less than that of $c_j$ in all other dimensions, i.e., 
      \begin{equation}
  c\neq \emptyset \wedge \forall k \in [0, d), \ci{c_i}{k} \le \ci{c_j}{k}  \wedge \exists k \in [0, d), \ci{c_i}{k} = \ci{c_j}{k}
  \end{equation} 
We use $c_i \sdom c_j$ to denote 
  that $c_i$ dominates or partially dominates $c_j$:
  \begin{equation}
    c_i \sdom c_j \iff c_i \dom c_j \lor c_i \pdom c_j
  \end{equation}
\end{definition}

By definition, a cell partially dominates itself, i. e.,  $c \sdom c$, and 
the ``$\sdom$'' relationship is transitive: 

\begin{lemma}\label{thm:inherit}
  If $c_i \sdom c_j$ and $c_j \sdom c_k$, then $c_i \sdom c_k$.
\end{lemma}
\begin{proof}
  Straightforward based on Definition~\ref{df:cd}.
\end{proof}
        
By cell domination, there are three types of cells in each layer. 
\begin{enumerate}
\item \emph{Dominated cells} -- cells that are dominated by some other cells, e.g., $\ls_4[14,4]$ in Fig.~\ref{fig:egcells} is dominated by $\ls_4[10, 1]$ which is non-empty (the dot in the cell represents a data point).
\item \emph{Irrelevant cells} -- cells that are neither dominated nor partially dominated, and do not dominate other cells, e.g., $\ls_4[2,1]$ in Fig.~\ref{fig:egcells}. 
These are empty cells with small column numbers.   
\item \emph{Candidate cells} -- cells that do not belong to the two types above, 
e.g., $\ls_4[10,1]$ in Fig.~\ref{fig:egcells}.
\end{enumerate}

No skyline points can be found from any cell $c_j$ dominated by another cell $c_i$, since points in $c_i$ must dominate  those in $c_j$. Thus, we can only find skyline points from  candidate cells.  
Next, we define candidate cells formally 
and bound the number of such cells. 

\section{Candidate Cells}
\label{sec:shrink_cs}

We first define candidate cells in Section~\ref{subsec:cc_definition}. Since we compute skyline points from candidate cells, the number of such cells determines the computation cost. 
We bound the number of candidate cells in Section~\ref{subsec:cc_bound}. 
We will detail our algorithms to compute candidate cells and hence the skyline points 
in the next section. 

\subsection{Defining Candidate Cells}\label{subsec:cc_definition}

\textbf{Key cells.} We first define a subset of the candidate cells -- the \emph{key cells}. Such cells form 
the basis of the set of candidate cells.  

\begin{definition}(Key cell)\label{df:keycell}
  We call a non-empty cell that is neither dominated nor partially dominated by any other cell a \emph{key cell}.
\end{definition}

We denote the set of all key cells in layer $i$ as $\ks_i$. For example, in Fig.~\ref{fig:egcells}, the cells marked by a ``$\ast$'' are the key cells. 
  
 Any cell that is either empty or dominated by a key cell (cf.~white cells in Fig.~\ref{fig:egcells}) cannot contain skyline points, and it is not a candidate cell. The remaining non-key cells each must be partially dominated by some key cell. 
We denote the set of cells partially dominated by a key cell $c$ but not  
dominated by other key cells as $\pdc(c)$.
\begin{equation}
  \label{eq:sss}
  \pdc(c) = \left\{c' \,\middle|\, c \pdom c' \land \nexists c'' \in \ks, c'' \dom c' \right\}.
\end{equation}
We call a cell in $\pdc(c)$ a \emph{partially dominated cell} of $c$. 
In Fig.~\ref{fig:egcells}, the gray cells in the same row or column of 
  a key cell are those partially dominated by the key cell. Such a cell may be partially dominated by multiple key cells, but this will not impact discussions below.

\textbf{Candidate cells.} 
The key cells and their partially dominated cells together form the set of candidate cells. 

\begin{definition}(Candidate cell)\label{df:canset}
  The set of candidate cells of the $i$-{th} layer, denoted by $\cs_i$, contains and only contains the key cells in $\ks_i$ and their partially dominated cells. Formally,
  \begin{equation}
    \label{eq:cst}
    \cs_i = \ks_i \cup \bigcup_{c\in \ks_i}\pdc(c). 
  \end{equation}
\end{definition}

\textbf{Auxiliary key cells.}  {Candidate cells discretize the 
 convex hull of $\mathcal{P}$ for skyline computation. To ensure no false dismissals, the candidate cells must cover the data space in each dimension. To derive the number of candidate cells, we use a set of \emph{auxiliary candidate cells} that covers each dimension. The number of such cells can be derived easily, and we can establish an one-on-one mapping between them and the candidate cells. We use \emph{auxiliary key cells} to simplify the description of  auxiliary candidate cells.}

To define auxiliary key cells, we first add $d$ \emph{auxiliary points} into the $d$-dimensional dataset $\ps$. The $i$-th auxiliary point, $\lambda_i$, satisfies $\lambda_i[i]=\underline{1}$ and $\lambda_i[j] = \underline{0}$ for all $0 \le j < d$, $j \ne i$. 
If $d=3$,  then $\lambda_0=(\underline{0}, \underline{0}, \underline{1})$,  $\lambda_1=(\underline{0}, \underline{1}, \underline{0})$, and $\lambda_2=(\underline{1}, \underline{0}, \underline{0})$. Here,  $\underline{1}$ (and $\underline{0}$) denotes a number infinitely close but \emph{less than} $1$ (and $0$). In Fig.~\ref{fig:egcells}, 
the ``$\ast$'' outside each grid layer denotes an auxiliary point. 

The auxiliary points are skyline points. 
However, they will not impact the skyline points of $\ps$.  
This is because our data points fall in $[0, 1)^d$. 
The auxiliary points will not dominate or be dominated by any point in $[0, 1)^d$ (including the origin), as they have coordinate $\underline{1}$  in some dimension and  
coordinate $\underline{0}$  in all other dimensions. 

The $d$ auxiliary points create $d$ additional key cells outside each grid layer. Such  cells are the \emph{auxiliary key cells}, e.g., the light-gray cells outside each layer in Fig.~\ref{fig:egcells} ($\Lambda_0$ and $\Lambda_1$ for Layer 4).  

\textbf{Auxiliary candidate cells.} The cells partially dominated by an  
auxiliary key cell are the auxiliary candidate cells. 

\begin{definition}(Auxiliary candidate cell)
  The set of \emph{auxiliary candidate cells} of the $i$-th layer, denoted by $\cps_i$, is formed by the cells partially dominated by some auxiliary key cell.
\end{definition}
    
In Fig.~\ref{fig:egcells}, the cells labeled by ``$\circ$'' are the auxiliary candidate cells of Layer 4. Such cells occupy the ``top'' column in each dimension.

\begin{lemma}\label{lm:cps}
  In the layer-$i$ grid of a $d$-dimensional space,
  \begin{equation}
    \label{eq:pcs}
    \cps_i = \left\{c \in \ls_i\, \middle| \, c[j] = 2^{i} -1, j \in [0, d)\right\}
  \end{equation}
\end{lemma}

\begin{proof}
  Let $\Lambda_j$ be the auxiliary key cell corresponding to auxiliary point $\lambda_j$ with coordinate $\underline{1}$ in dimension $j$ and $\underline{0}$ in the other dimensions.  By Equation~\ref{eq:cell_column_number}, $\Lambda_j[j] = 2^{i} - 1$ and 
  $\Lambda_j[k] = -1$, $\forall k \ne\lb j, 0 \le k < d$.
  For example, in Fig.~\ref{fig:egcells}, the bottom-right light-gray cell of Layer 4
    is $\Lambda_0 = \ls_4[-1, 15]$.
  By Definition~\ref{df:cd}, 
  any cell $c \ne \Lambda_j$ and $c[j] = \Lambda_j[j]$ is partially dominated by $\Lambda_j$, i.e., $\pdc(\Lambda_j) = \left\{c \,\middle|\, c[j] = 2^{i} -  1\right\}$. 
  In Fig.~\ref{fig:egcells}, the cells in the same column as $\Lambda_0$ form 
    $\pdc(\Lambda_j)$.
  Combining $\pdc(\Lambda_j)$ for all $j \in [0, d)$, we obtain Equation~\ref{eq:pcs}.
\end{proof}

\subsection{Bounding the Number of Candidate Cells}\label{subsec:cc_bound}

We show the number of candidate cells to be a function of $i$ and $d$, which is independent of the dataset size. 
This is done by a one-to-one mapping between 
the  candidate cells and the auxiliary candidate cells, the number of which 
can be derived from Equation~\ref{eq:pcs}.

\begin{theorem}\label{cell_in_C}
  In Layer $i$, 
  there is a bijection between the set of candidate cells $\cs_i$ and the set of auxiliary candidate cells $\cps_i$.
\end{theorem}
\begin{proof} 
  We first construct a mapping from a cell $c\in \cs_i$ to a cell $ca \in \cps_i$ and then show that it is one-to-one, i.e., a bijection.

  \underline{Mapping construction.} 
Given a candidate cell $c \in \cs_i$, we use $K_c \subset \cs_i$ to denote the set of key cells that partially dominate $c$ (a key cell is also a candidate cell, and it partially dominates itself). 
  
  We define function $\omega(c, ck)$ to return   
  the highest dimension where $c$ and $ck \in K_c$ have the same column index. 
  Function $\Omega(c)$ further returns 
  the minimal value of $\omega(c, ck)$ for all $ck\in K_c$: 
  \begin{align}
    \omega(c, ck) &= arg \max_j \big(c[j] = ck[j]\big),\label{somega}\\
    \Omega(c) &= \min_{ck\in K_c}\omega(c, ck).\label{bomega}
  \end{align}
  In Layer 4, Fig.~\ref{fig:egcells}, 
   $c=\ls_4[4,6]$ is partially dominated by $ck_1 = \ls_4[1,6]$ and $ck_2 = \ls_4[4,4]$. Since $c[0] = ck_1[0] = 6$, $\omega(c, ck_1) = 0$. Similarly, 
  $\omega(c, ck_2) = 1$, and $\Omega(c) = 0$.

  We define a mapping $\Psi(c)$ from cell  $c \in \cs_i$ to a cell $ca \in \cps_i$:  
  \begin{equation}\label{eq:dfpi}
    \Psi(c)= ca,\;\;\;\;s.t.
    \begin{cases}
      ca[j] = 2^i-1,\;\;&j=\Omega(c),\\
      ca[j] = c[j],\;\;&\text{otherwise}.
    \end{cases}
  \end{equation}
  For cell $c=\ls_4[4,6]$ in Fig.~\ref{fig:egcells}, it will be mapped to $ca = \ls_3[4,15]$.

  We show that $\Psi(\cdot)$ is a bijection below. 
  
  \underline{Injection proof.} 
First, we show that $\Psi(\cdot)$ is injective, i.e., 
  for $c \in \cs_i$, $\Psi(c)\in \cps_i$, and for another $c' \in  \cs_i$ ($c \ne c'$), $\Psi(c) \ne \Psi(c')$.

  By Equation~\ref{eq:dfpi}, for $c \in \cs_i$, $ca = \Psi(c)$ must have 
   a dimension $j$ where $ca[j]$ is $2^i-1$. By Lemma~\ref{lm:cps}, 
  $ca$ must be in $\cps_i$.

  For two cells $c\ne c'$, there must be some dimension(s) where the column indices of $c$ and $c'$ differ. There are three cases:
  \begin{itemize}[leftmargin=*]
  \item Case 1: Cells $c$ and $c'$ have different column indices in more than two dimensions. 
    Since $\Psi(c)$ and $\Psi(c')$ only change the column indices of $c$ and $c'$ 
    in at most one dimension, respectively, their column indices differ in at lease one dimension, i.e., $\Psi(c) \ne \Psi(c')$.
    
  \item Case 2:  Cells $c$ and $c'$ have different column indices in two dimensions.
    Let $ca = \Psi(c)$ and $ca' = \Psi(c')$. Let $j$ be a  dimension where $c[j] \ne c'[j]$. If $j \ne \Omega(c)$ and $j \ne \Omega(c')$, by Equation~\ref{eq:dfpi},  we have $ca[j] = c[j] \ne c'[j] = ca'[j]$. 
    Thus, we only need to consider the case where $j = \Omega(c)$ or $\Omega(c')$, i.e., 
    $c$ and $c'$ have different column indices in dimensions $\Omega(c)$ and $\Omega(c')$. 
    \begin{itemize}[leftmargin=*]
    \item Case 2a:
      If $j = \Omega(c) = \Omega(c')$, there must be another dimension $j' \ne j$ where $c[j] \ne c'[j]$.  Then, $ca[j'] \ne ca'[j]$ and $ca \ne ca'$. 
      
    \item Case 2b:   If $j = \Omega(c) \ne \Omega(c')$, $ca[j] = 2^i-1$. For $ca'[j]$ to be the same as $ca[j]$, $c'[j] = 2^i-1$. Cell $c'$ is thus partially dominated by auxiliary key cell $\Lambda_j$ which contains $\lambda_j$. By Equation~\ref{bomega}, we have $\Omega(c') \le j$. Further, since $j \ne \Omega(c')$, we have $\Omega(c') < j = \Omega(c)$.
      Now consider the other dimension $j' = \Omega(c') \ne \Omega(c)$ where $c$ and $c'$ have different column indices. Following the same argument, we have $\Omega(c) < j' = \Omega(c')$. Since $\Omega(c)$ and $\Omega(c')$ cannot be less than each other at the same time, we derive a contradiction. Thus, $ca$ and $ca'$ cannot be the same.  
      
    \end{itemize}

  \item  Case 3: Cells $c$ and $c'$ have different column indices in one dimension.
    Let $j$ be this dimension, i.e., $c[j] \ne c'[j]$. 
    Let $ca = \Psi(c)$ and $ca' = \Psi(c')$.
    There are again three sub-cases:
    \begin{itemize}
    \item Case 3a: $j\ne \Omega(c)$ and $j\ne \Omega(c')$. Then, 
      $ca[j] = c[j] \ne c'[j] = ca'[j]$. Thus, $\Psi(c) \ne \Psi(c')$.

    \item Case 3b: $j = \Omega(c)$ or $j = \Omega(c')$ while $ \Omega(c) \ne \Omega(c')$.
      We consider  $j = \Omega(c) \ne \Omega(c')$. The case where $j = \Omega(c') \ne \Omega(c)$ is symmetric and is omitted for conciseness. We show that $c'[j]\ne 2^i -1$ and thus 
      $ca[j] = 2^i -1 \ne ca'[j]$  by contradiction.
      Suppose $c'[j]= 2^i -1$. Then, $c'$ is partially dominated by $\Lambda_j$. By Equation~\ref{bomega} and $\Omega(c') \ne j$, we have $\Omega(c') < j$. Since $c[\Omega(c')]=c'[\Omega(c')]$ (recall that $c[k]=c'[k]$ when $k\ne j, 0\le k < d$), $c$ is partially dominated by the same key cell $ck$ with $ck[\Omega(c')] = c'[\Omega(c')]$. By Equation~\ref{bomega}, $\Omega(c) \le \Omega(c')$. Therefore, $\Omega(c) < j$,  and we have a contradiction.
      
    \item Case 3c: $j=\Omega(c) = \Omega(c')$. We show that this is infeasible  by contradiction. Suppose $j = \Omega(c) = \Omega(c')$.  Let $c[j] < c'[j]$ (the case where $c'[j] < c[j]$ is the same and omitted). For the key cell $ck$ that yields 
      $\Omega(c)$, by Equation~\ref{somega},  
      we have $ck[k] \le c[k] = c'[k]$ ($\forall 0 \le k < j$), $ck[j] = c[j] < c'[j]$, 
      and $ck[k] < c[k] = c'[k]$ ($\forall j < k < d$).
      Thus,  $ck$ also partially dominates $c'$, and $\omega(c', ck) < j$. 
      This contradicts the fact that $\Omega(c') = j$.
    \end{itemize}
  \end{itemize}
  
  \underline{Surjection proof.} Next, we show that $\Psi(\cdot)$ is surjective, i.e., for each $ca \in \cps_i$, there exist $c\in \cs_i$ such that $\Psi(c) = ca$. 
  We define a function $\Theta(ca)$ to map from $\cps_i$ to $\cs_i$: 
  \begin{equation}
    \label{eq:mapback}
    \Theta(ca) = c,\;\;\;\;
    \begin{cases}
      c[j] = \Phi(ca),\;\;&j=\gamma(ca),\\
      c[j] = ca[j],\;\;&\text{otherwise}.
    \end{cases}
  \end{equation}
  Here, $ca\in \cps_i$, and $\gamma(ca)$ is the smallest dimension where the column 
  index of $ca$ is $2^i-1$: 
  \begin{equation}
    \label{eq:rho}
    \gamma(ca) = arg \min_j \big(ca[j] = 2^i-1\big)
  \end{equation}
  
  We define $\Phi(ca)$ as: 
  \begin{equation}
    \label{eq:upsilon}
    \Phi(ca) = \min_{ck\in \ks_i}\{ck[\gamma(ca)]\}, s.t.  
    \begin{cases}
      ck[j] \le ca[j],&j\le\gamma(ca),\\
      ck[j] < ca[j],&j>\gamma(ca).
    \end{cases}
  \end{equation}
  In Fig.~\ref{fig:egcells}, for $ca = \ls_4[4, 15]$, $\gamma(ca) = 0$. Key cells 
  $ck_1 = \ls_4[1,6]$ and $ck_2 = \ls_4[0,10]$ 
  both satisfy $ck[j] < ca[j]$ when $j >\gamma(ca)$, i.e., for $j =1$, $ck_1[1] = 1$ and  $ck_2$[1] = 0 are both smaller than $ca[1] = 4$. 
  Thus, $\Phi(ca) = \min\{ck_1[\gamma(ca)], ck_2[\gamma(ca)]\} = ck_1[0] = 6$, and $c = \Theta(ca) = \ls_4[4, 6]$. 
  
  Next, we prove that $\Theta(ca)\in \cs_i$ and $\Psi(\Theta(ca)) = ca$.

  \begin{itemize}[leftmargin=*]
  \item $\Theta(ca)\in \cs_i$.  For any $ca\in \cps_i$, there is at least one $ck \in \ks_i$ satisfying Equation~\ref{eq:upsilon}, i.e.,  the auxiliary key cell $c = \Lambda_{\gamma(ca)}$\footnote{The auxiliary key cells are considered to be in the set of key cells $\ks_i$ in the proof. They map to themselves in the bijection $\Psi(\cdot)$.}.
    We have $c[\gamma(ca)]=2^i - 1 = ca[\gamma(ca)]$ and $c[j] = -1 < ca[j]$ for any $0\le j < d$ and $j \ne \gamma(ca)$. 
    Among all key cells satisfying Equation~\ref{eq:upsilon}, let $ck$ be the one with the minimum column index in 
    dimension $\gamma(ca)$. By Equation~\ref{eq:mapback}, 
    $ck$ partially dominates $ \Theta(ca)$, i.e., $ck \pdom \Theta(ca)$. 
    In Fig.~\ref{fig:egcells}, 
    for $ca = \ls_4[4, 15]$, $ck = \ls_4[1,6] \pdom \Theta(ca) = \ls_4[4, 6]$.
    Meanwhile, $\Theta(ca)$ is not dominated by any key cell. 
    Otherwise, let such a key cell be $ck'$. Then, $ck'$ satisfies Equation~\ref{eq:upsilon}, and $ck'[\gamma(ca)] < ck[\gamma(ca)]$. This contradicts the fact that $ck$ has the minimum column index in dimension $\gamma(ca)$. Since $\Theta(ca)$ is partially dominated by a key cell but not dominated by any key cell, it must be a candidate cell, i.e.,  $\Theta(ca) \in \cs_i$. 
    
  \item $\Psi(\Theta(ca)) = ca$. 
    Let $c = \Theta(ca)$. 
    Based on Equations~\ref{eq:dfpi} and~\ref{eq:mapback}, we only need to show $\Omega(c) = \gamma(ca)$ to prove $\Psi(c) = ca$. This is because   
    $\Psi(c)$ and $\Theta(ca)$ only change the column indices of  $c$ and $ca$ in one dimension, i.e., dimension $\Omega(c)$ and $\gamma(ca)$, respectively. 
    Let $ck$ be the key cell that satisfies the conditions in Equation~\ref{eq:upsilon} 
    and has the minimum column index in dimension $\gamma(ca)$. 
    Then, $ck[j]\le ca[j]$ when $j<\gamma(ca)$, $ck[j] < ca[j]$ when $j > \gamma(ca)$, and $ck[\gamma(ca)] = \Theta(ca)[\gamma(ca)]$. Thus, $ck\pdom \Theta(ca) = c$. 
    Next, we show that $ck$ is the key cell that yields $\Omega(c)$ in Equation~\ref{bomega}.
    Since $ck[j] < ca[j]$ when $j > \gamma(ca)$, by Equation~\ref{somega}, we have $\omega(c, ck)=\gamma(ca)$. Assume another key cell $ck' \pdom c$, and $\omega(c, ck')$ < $\omega(c, ck)$. Then, $ck'$ also satisfies the conditions in Equation~\ref{eq:upsilon}, and $ck'[\gamma(ca)] < ck[\gamma(ca)]$. This contradicts the fact that $ck$ has the minimal column index in dimension $\gamma(ca)$. Therefore,  $\omega(c, ck)$ must be the smallest among all $ck\in \ks_c$, i.e., $ck$ yields $\Omega(c)$, and $\Omega(c) = \gamma(ca)$.  
    This completes the proof.
  \end{itemize} 
\end{proof}

\textbf{Bounding candidate cells of a layer.} 
Given Lemma~\ref{lm:cps} and Theorem~\ref{cell_in_C}, 
we  bound the number of candidate cells as follows. 
\begin{corollary}\label{thm:ratio}
  In a $d$-dimension space, the number of candidate cells in Layer $i$, denoted by $|\cs_i|$, is computed as: 
  \begin{equation}
    \label{eq:candnum}
    |\cs_i|=\sum_{j=0}^{d-1}(2^{i}-1)^j \cdot 2^{i(d-1-j)}
  \end{equation}
\end{corollary}

\begin{proof}
  By Theorem~\ref{cell_in_C}, the number of candidate cells is the same as the
  number of auxiliary candidate cells.  
  By Lemma~\ref{lm:cps},  $\cps_i = \left\{c \in \ls_i\, \middle| \, c[j] = 2^{i} -1, j \in [0, d)\right\}$. We derive the number of cells in $\cps_i$ for each $j \in [0, d)$ to derive 
  $|\cps_i|$ and hence $|\cs_i|$.
  \begin{itemize}[leftmargin=*]
  \item For $j = 0$, we have $c[0] = 2^{i}-1$. The number of such cells is: 
    $$2^{i (d-1)} = (2^{i}-1)^0 \cdot 2^{i(d-1-0)}\vspace{-1mm}\vspace{-1mm}$$
    These cells form a slice of a $[2^{i}]^d$ grid, e.g., a column 
    in a two-dimensional grid (cf. the right-most column of Layer 4 in Fig.~\ref{fig:egcells}).  
    
  \item For $j = 1$, we have $c[1] = 2^{i}-1$, and $c[0] \ne  2^{i}-1$ to avoid counting the same cells twice. 
The number of such cells is: 
    $$(2^{i}-1)\cdot 2^{i (d-2)}= (2^{i}-1)^1 \cdot 2^{i(d-1-1)}\vspace{-1mm}$$
 In dimension-0, there are $2^{i}-1$ possible column indices for these cells (one column less due to $c[0] \ne  2^{i}-1$); in dimension-1, there is just one possible column index ($c[1] = 2^{i}-1$);  and in each of the other $d-2$ dimensions, there are  $2^{i}$  possible column indices (cf. the top row without the top-right cell of Layer 4 in Fig.~\ref{fig:egcells}).
    
  \item In general, for $j = k$, we have $c[k] = 2^{i}-1 
    \land c[0] \ne 2^{i}-1 \land c[1] \ne 2^{i}-1 \land \cdots\land c[k-1] \ne 2^{i}-1$. 
    The number of such cells is: 
    $$(2^{i}-1)^k\cdot 2^{i (d-1-k)}\vspace{-1mm}$$
  \end{itemize}
  Summing up the numbers for $j \in [0, d-1]$ yields Equation~\ref{eq:candnum}.
\end{proof}

In Fig.~\ref{fig:egcells}, the numbers of candidate cells in Layers $0$ to $4$ when $d = 2$ are 1, 3, 7, 15 and 31, which conform to the corollary.

\textbf{Bounding candidate cells across layers.} 
The candidate cells in different layers further satisfy the following two corollaries, which  
enable their efficient computation.   
\begin{corollary}\label{crl:decrease}
  Given $i > j$, the volume (or area if $d=2$) covered by the cells in $\cs_i$ must be smaller than that by the cells in $\cs_j$.
\end{corollary}
\begin{proof}
Intuitively, this is because candidate cells of a higher layer are all covered by 
those of a lower layer (cf. Fig.~\ref{fig:egcells}). 

  Recall that the number of candidate cells in Layer $i$ is 
  $\sum_{k=0}^{d-1}(2^{i}-1)^k \cdot 2^{i(d-1-k)}$. This is the sum of a geometric sequence, which adds up to  $2^{i \cdot d}-(2^{i}-1)^d$. In this layer, the data space is partitioned into $2^{i \cdot d}$ cells, where each cell has volume (or area) $1/2^{i  \cdot  d}$. 
  Thus,  the candidate cells in $\cs_i$ cover a volume (or area) of 
   $\mathcal{V}_i = \left(2^{i \cdot d}-(2^{i}-1)^d\right)/2^{i \cdot d} = 1 - (2^{i}-1)^d/2^{i \cdot d}$. 
Similarly, we can write out the volume (or area) $\mathcal{V}_j$  covered by the cells in $\cs_j$ (by replacing every $i$ with $j$).  
  By basic arithmetic, we can show $\mathcal{V}_i - \mathcal{V}_j < 0$.  Thus, the volume covered by the cells in $\cs_i$ is smaller than that by the cells in $\cs_j$. We omit the detailed  calculation due to space limit. 
%
\end{proof}

The following corollary suggests that a key cell in Layer $i$ must yield at least a key cell in Layer $i+1$.


\begin{corollary}\label{thm:1in4}
  Given a key cell $ck$ in Layer $i$, let $\spt(ck)$ be the 
  set of cells resulted from partitioning $ck$ in Layer $i+1$. 
  There exists at least a key cell in $\spt(ck)$, i.e., $\exists ck' \in \spt(ck) \land ck' \in \ks_{i+1}$.
\end{corollary}

\begin{proof}
Every cell 
  in Layer $i$, including a key cell $ck$, is partitioned into $2^d$ cells in Layer $i+1$, e.g., a cell in Layer 0 in Fig.~\ref{fig:egcells} is partitioned into $2^2=4$ cells in Layer 1. Thus, 
  $\spt(ck) \ne \emptyset$. 

  Recall that a key cell $ck$ is non-empty (i.e., containing data points), and there must be non-empty cells in $\spt(ck)$. Among such cells, there must be a cell  $ck'$ 
  that is not dominated by the other cells in $\spt(ck)$ (the cells cannot all dominate each other). 
  
  We also have that $ck'$ is not 
   dominated or partially dominated by a cell $c' \in \spt(c)$ that is created by partitioning any other cell $c$ ($c \ne ck$) in Layer $i$. 
   Otherwise, $c'[k] \le ck'[k], \forall k \in [0, d)$. This means  
  $c[k] \le ck[k], \forall k \in [0, d)$, i.e.,  $c \sdom ck$, which contradicts the fact that $ck$ is a key cell. This completes the proof. 
\end{proof}

In Fig.~\ref{fig:egcells}, key cell $\ls_1[0,0]$ (marked by``$\ast$'') in 
Layer~1 yields 
key cell $\ls_2[0, 1]$ in Layer 2, which yields key cell $\ls_3[0,3]$ in Layer 3. 

Next, we detail our skyline algorithms based on candidate cells.

\section{Query Processing}
\label{sec:ps}

We first present our overall algorithm named \textbf{SkyCell}. We will then detail a key sub-procedure named \texttt{ShrinkKeyCells}  in Sections~\ref{sec:sequentialc} and~\ref{sec:pave}, for its  
sequential and parallel design, respectively. 
  
\textbf{SkyCell algorithm.} 
As summarized in Algorithm~\ref{alg:sq}, SkyCell 
first computes a $\rho$-layer ($\rho$ is detailed next) grid partitioning over dataset $\ps$ (Line 1).  We store the points in an array and sort them according to the Layer-$\rho$ cells to which they belong. Any cell ordering can be used, e.g., the Z-order. We just require points from the same cell to occupy a consecutive segment of the array. Then, for each Layer-$\rho$ cell, we record the starting and ending array indices of the points in the cell. An empty cell has the same starting and ending array indices. 
This constructs $\ls_\rho$ of our grid structure. 

\begin{algorithm}
  \begin{small}
    \caption{SkyCell}\label{alg:sq}

    \SetAlgoNoLine
    \SetKwInOut{Input}{input}
    \SetKwInOut{Output}{output}
    \SetKwFunction{shrink}{ShrinkKeyCells}
    \SetKwFunction{refine}{RefineSkyline}

    \Input{\ Dataset $\ps$}
    \Output{\ Skyline set \sss}
    Compute $\ts_\rho$ to $\ts_0$ from $\ps$\;
    $\rs_0 \leftarrow \ts_0[0,\ldots,0]$\label{ag:1:initial}\;
    \For{$i=0$ \KwTo $\rho - 1$}{
      $\rs_{i+1} \leftarrow $\shrink{$\ps$, $i$, $\rs_i$, $\ts_{i+1}$}\label{ag:1:call}\;
    }
    \Return \refine{$\ps$, $\rs_\rho$}\;
  \end{small}
\end{algorithm}
  
We construct $\ls_{\rho-1}$ from $\ls_\rho$. For each cell $c \in \ls_{\rho-1}$, we record whether it is non-empty (encloses data points), which will be used for key cell testing later. This is done by a simple scan over the starting and ending array indices of the cells in $\spt(c)$. Similarly, we construct the other layers from $\ls_{\rho-2}$ back to $\ls_0$ (Line~1).


Then, we compute a set $\rs_i$ of cells of interest for each Layer $i$ based on Corollary~\ref{thm:1in4} with a sub-procedure named \texttt{ShrinkKeyCells} (Lines 2 to 4, detailed later). For our sequential algorithm, $\rs_i$ contains key cells ($\rs_i=\ks_i$). For our parallel algorithm, $\rs_i$ contains key cells and candidate cells ($\rs_i=\{\ks_i, \cs_i\}$).  Here, $\ls_0$ has only one cell (i.e., the entire data space), which is used as $\ks_0$ and $\cs_0$. 

When $\rs_\rho$ is computed, $\ks_\rho$ is also computed. We use 
 $\ks_\rho$ to compute $\cs_\rho$ following a procedure similar to \texttt{ShrinkKeyCells}, which also computes candidate cells from key cells (details omitted for succinctness). We then compute skyline points from each candidate cell in $\cs_\rho$ and return them as the result. As points from different candidate cells do not  dominate each other, the 
candidate cells are processed in parallel (for parallel SkyCell). 
We use the \emph{sort-first skyline} (SFS)~\cite{chomicki2003skyline} algorithm to compute 
the skyline points in each cell, while other algorithms may also apply.  Sub-procedure \texttt{RefineSkyline} summarizes these steps (Line~5). 


\textbf{Partition ratio $\rho$.} 
Parameter $\rho$  balances the workload of 
key cell computation in multiple layers and the workload of candidate cell computation 
and skyline point checking in Layer $\rho$.  
We call this parameter the \emph{partition ratio} and will evaluate its impact empirically. 

\subsection{Sequential Key Cell Shrinking}
\label{sec:sequentialc}
We detail sequential $\mathtt{ShrinkKeyCells}$ in this subsection. 
We first show that key cells in $\ks_{i+1}$ must come from partitioning  
 candidate cells in $\cs_{i}$. Then, we show how to enumerate the candidate cells 
in $\cs_{i}$ from the key cells in $\ks_{i}$. We generate the key cells in $\ks_{i+1}$
during this process, which yields sequential $\mathtt{ShrinkKeyCells}$.  

\textbf{Relationship between $\ks_{i+1}$ and $\cs_i$.}  
We show that a key cell $ck_{i+1} \in \ks_{i+1}$ must be from partitioning a candidate cell $c_i \in \cs_i$. 

\begin{corollary}\label{thm:kiplus1_to_ki}
  Given a key cell $ck_{i+1} \in \ks_{i+1}$, 
  there exists a  candidate cell $c_i \in \cs_i$, such that $ck_{i+1} \in \spt(c_i)$, i.e., 
  \begin{equation}
    \label{eq:findkj1}
    \ks_{i+1} \subset \spt(\cs_i) = \spt(\ks_i \cup \bigcup_{ck\in\ks_i}\pdc(ck)).
  \end{equation}
\end{corollary}

\begin{proof}
  We prove by contradictory. 
  Suppose $ck_{i+1}$ is created from a Layer-$i$ cell $c_i \notin \cs_i$.  
  Since $c_i$ is not a candidate cell, it must be either empty or 
  dominated by some key cell $ck_i \in \ks_i$. 
  
\begin{enumerate}[leftmargin=*]
\item If $c_i$ is empty, $ck_{i+1}$ must also be empty and not a key cell. 

\item If $c_i$ is dominated by $ck_i \in \ks_i$, 
  based on Corollary~\ref{thm:1in4}, $ck_i$ must yield at least a key cell $ck'_{i+1} \in \ks_{i+1}$. 
  Since  $c_i$ is dominated by $ck_i$,  any cell in $\spt(c_i)$ is also dominated by every cell in $\spt(ck_i)$. Thus, $ck_{i+1} \in \spt(c_i)$ must be dominated by $ck'_{i+1} \in \spt(ck_{i})$, and hence is not a key cell. 
  \end{enumerate}
\end{proof}

In each grid layer in Fig.~\ref{fig:egcells}, we can see that the key cells (marked by ``$\ast$''
correspond to candidate cells of the previous layer. 

\begin{figure}[h]
  \centering
  \includegraphics[width=8.5cm]{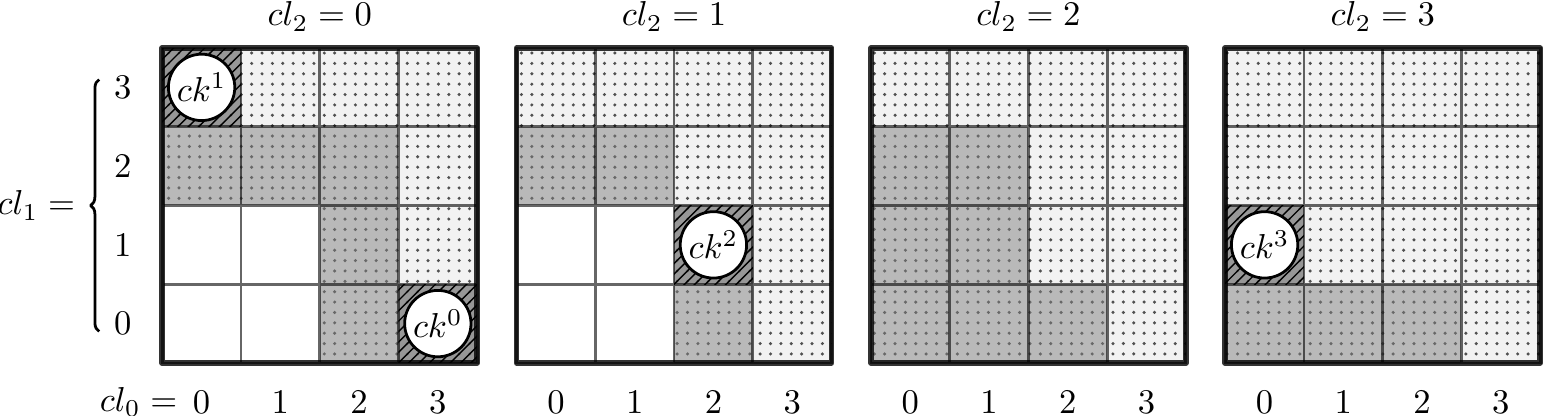}
  \caption{Example of candidate cell enumeration}
  \label{fig:cce} 
\end{figure}

\textbf{Enumerating the candidate cells in $\cs_i$.} 
All cells in a layer can be enumerated by their column indices. 
By carefully controlling the enumeration process, we can 
also enumerate all candidate cells in a layer by their column indices. 
We use Fig.~\ref{fig:cce} to help illustrate our enumeration procedure. 
The figure shows the Layer-2 grid of a 3-dimensional space, 
where dimension 2 (i.e., dimension $d-1$ which is the most significant dimension) 
is represented by the four grids (think of them as stacking from $cl_2=0$ to $cl_2 = 3$). 
The dotted cells are partitioned from the Layer 1 candidate cells (i.e., all cells in $\ls_1$ are candidate cells except $\ls_1[0, 0, 0]$), assuming that there are just the three auxiliary key cells $\ls_1[-1,-1,1]$, $\ls_1[-1,1,-1]$ and $\ls_1[1,-1,-1]$ in Layer 1 and no other key cells. 


The dotted cells in Fig.~\ref{fig:cce}  form $\spt(\ks_1)$. We enumerate them to find key cells in Layer 2, $\ks_2$. This is done by the order of column indices from dimensions $d-1$ to 0, i.e., enumerating $[cl_2, cl_1, cl_0]$ from $[0, 0, 0]$ to $[3, 3, 3]$. First, consider $cl_2=0$. At $cl_1=0$, suppose $\ls_2[0,0,3]$ is found to be the first non-empty cell. This must be a key cell in $\ks_2$ by definition, denoted by $ck^0$. Now $cl_0= 3$. We increase  $cl_1$ by one ($cl_1=1$) and reset $cl_0$. We check up to $cl_0 = 2$, because 
a key cell $ck^0$ has been found at $cl_0 = 3$, which will partially dominate $\ls_2[0, 1, 3]$. 
Repeating this procedure, we  enumerate the cells for $cl_1 = 2$ ($cl_0$ also up to 2). 
There is no non-empty cell found, and we move on to $cl_1 = 3$. 
 Suppose that we find another non-empty cell, i.e., a key cell $ck^1 = \ls_2[0, 3, 0]$. We do not need to enumerate for  $cl_0 > 0$, because now there is a key cell at $cl_0 = 0$.  

Now we move onto $cl_2=1$. We enumerate $[cl_1, cl_0]$ again. Note that $cl_0$ only needs to reach 2, because of key cell $ck^0 = \ls_2[0,0,3]$. We find a third key cell $ck^2 = \ls_2[1,1,2]$. This further limits $cl_0$ to be less than 2. The process repeats, and there is no key cell for $cl_2=2$. At $cl_2=3$, there is a fourth key cell $ck^3 = \ls_2[2,1,0]$. The enumeration terminates because $ck^3$ limits $cl_0$ to be less than 0. 



The enumeration above collects all non-empty cells that are not dominated by other cells, i.e., key cells in $\ks_2$. They also prune part of $\spt(\ks_1)$ from being enumerated (only the gray cells in Fig.~\ref{fig:cce} have been enumerated), which reduce the computation costs. 


\textbf{Sequential \texttt{ShrinkKeyCells}.} Our sequential \texttt{ShrinkKeyCells} follows the idea above to go through the cells in $\spt(\cs_i)$ to generate the key cells in $\ks_{i+1}$. As summarized in Algorithm~\ref{alg:sc}, \texttt{ShrinkKeyCells} enumerates all column index combinations for dimensions $d-1$ to $1$ but considers the column index in dimension 0 ($cl_0$) separately. The value range of $cl_0$ is constrained by the start index of the candidate cells in $\cs_i$ and the key cells in $\ks_{i+1}$ found. This enables pruning the enumeration. 

\begin{algorithm}
  \begin{small}
    \caption{ShrinkKeyCells (Sequential)}\label{alg:sc}

    \SetAlgoNoLine
    \SetKwInOut{Input}{input}
    \SetKwInOut{Output}{output}
    \SetKwFunction{cost}{CostLine}
    \SetKwFunction{adjust}{Adjust}
    \SetKwFunction{append}{Append}
    \SetKwFunction{MinGS}{MinGS}
    \SetKwFunction{MinGE}{MinGE}
    \SetKwFor{enumerate}{enumerate}{do}{end}

    \Input{\ Current layer number $i$,
      key cells $\ks_i$, cells $\ts_{i + 1}$}
    \Output{\ Key cells  $\ks_{i+1}$}
    $\ks_{i+1}=\emptyset,\; j=0$\label{sc:initial}\;
    \For{$cl_{d-1}=-2$ \KwTo $2^i-1$} { \label{sc:outer}
      $\cdots$\tcc*{$\bbii \text{ denotes } [cl_{d-1},\ldots,cl_1]$}
      \For{$cl_{1}=-2$ \KwTo $2^i-1$} {
        \If {$\bbii=\big[2\cdot ck_i^{j}[d-1], 2\cdot ck_i^{j}[d-2], \ldots, 2\cdot ck_i^{j}[1]\big]$}{\label{sc:setI}
          $\gs_s[\bbii] \leftarrow 2\cdot ck_i^{j}[0], j\leftarrow  j+1$ \label{sc:css}\;
        }
        $\gs_s[\bbii]\leftarrow \MinGS{\bbii}$,
        $\gs_e[\bbii]\leftarrow \MinGE{\bbii}$\label{sc:adjust}\;
        \If {$ck = \ls_{i+1}[\bbii,-1]$  \text{\emph{ or }}  $\ls_{i+1}[\bbii,2^{i}-1]$ \text{\emph{ is an auxiliary key cell }} } {
      		$\ks_{i+1} \xleftarrow{\text{append}}  ck$, \textbf{continue}\label{sc:appendAKC}\; 
      	}
        \If {\bbii \text{\emph{does not contain negative indices}}}{\label{sc:valid}
          \For{$cl_0= \gs_s[\bbii]$ \KwTo $\gs_e[\bbii]$}{ \label{sc:cost}
            \If {$\ts_{i + 1}[\bbii, cl_0]\emph{ is not empty}$}{ \label{sc:if}
              $\gs_e[\bbii] \leftarrow cl_0-1$\label{sc:appcse}\;
              \If {$\texttt{NotPartiallyDomed}(\ts_{i+1}[\bbii,cl_0])$}{
              $\ks_{i+1} \xleftarrow{\text{append}}  \ts_{i+1}[\bbii, cl_0 ]$\label{sc:append}\;            
              }
              \textbf{break}\label{sc:end}\;
            }
          }
        }
      }
    }
    \Return $\ks_{i+1}$\;
  \end{small}
\end{algorithm}

We use $\bbii$ to denote a column index combination for dimensions $d-1$ to $1$, i.e., $\ls_{i+1}[\bbii, cl_0]$ is the index of an enumerated cell. The enumeration of each dimension in 
$\bbii$ starts at $-2$ (Lines 2 to 4). This is because the auxiliary key cells have indices $-1$ in Layer $i$, which doubles to $-2$ in Layer $i+1$. Not that for a cell $c \in \ls_i$ with column index $cl_j$ in dimension $j$,  $\spt(c)$ contains cells with column indices starting at $2 cl_j$ in dimension $j$.

For each \bbii, we compute $\gs_s[\bbii]$ and $\gs_e[\bbii]$ to bound the value of $cl_0$ to be enumerated (Lines~\ref{sc:setI} to \ref{sc:adjust}, detailed below).
We test if \bbii can form an auxiliary key cell. If so, we add it to $\ks_{i+1}$ and move 
onto the next \bbii combination (Line~\ref{sc:appendAKC}). 
 If not, and \bbii does not contain negative indices,  
we enumerate $cl_0$  to check if $\ls_{i+1}[\bbii, cl_0]$ is non-empty (Lines~\ref{sc:valid} to \ref{sc:if}). Once we find a non-empty cell $c$, 
$\gs_e[\bbii]$ is updated to the current $cl_0-1$ (Line~\ref{sc:appcse}). 
We check whether $c$ is partially dominated by an auxiliary key cell in $\ks_{i+1}$ 
(by \texttt{NotPartiallyDomed}). If not, then $c$ is a key cell, and we add it to $\ks_{i+1}$ (Line~\ref{sc:append})\footnote{In Fig.~\ref{fig:cce}, we added $ck^0$, $ck^1$, and $ck^3$ for ease of illustration. In actual implementation, the auxiliary key cells that partially dominate them are added instead.}. We then move on to the next \bbii combination (Line ~\ref{sc:end}).

\textbf{Bounding $cl_0$.} 
Range $[\gs_s[\bbii], \gs_e[\bbii] ]$ bounds  $cl_0$ given $\bbii$. We store 
$\gs_s[\bbii]$ and $\gs_s[\bbii]$ each in a $(d-1)$-dimensional table (because \bbii has $d-1$ dimensions). Given \bbii, we first test whether it is now at a slice that overlaps a key cell $ck_i^j \in \ks_i$, i.e., $\bbii=\big[2\cdot ck_i^{j}[d-1], 2\cdot ck_i^{j}[d-2], \ldots, 2\cdot ck_i^{j}[1]\big]$ (recall that column indices for cells in $\spt(c)$ start from 
$[2\cdot c[d-1],  2\cdot c[d-2], \ldots, 2\cdot c[0]]$). 
If so, $\gs_s[\bbii]$ should start from $2 \cdot ck_i^j[0]$. This is because cells at $[\bbii, cl_0]$ where $cl_0 < 2 \cdot ck_i^j[0]$ must be empty. Otherwise, there will be a non-empty cell 
$ck' \in \ks_i$ with $ck'[k] = ck_i^j[0]$ ($\forall 0 < k \le d-1$) and $ck'[0] < ck_i^j[0]$, which partially dominates $ck_i^j$.  We also increase $j$ by 1 such that later \bbii's can check against the next $ck_i^j$  (Line~\ref{sc:css}, note that key cells are added to $\ks_i$ in the  index enumeration order). 


We further adjust $\gs_s[\bbii]$ and $\gs_e[\bbii]$ by  $\gs_s$ and $\gs_e$ of previously seen column index combinations, because key cells yielded by those combinations may dominate or partially dominate cells generated by \bbii. Such dominated cells can be pruned.  We check $d-1$ previous index combinations, where each combination differs from $\bbii$ in one  dimension. The $k$-th ($0 < k \le d-1$) previous combination, $\bbii_k$, satisfies $\bbii_k[k] = \bbii[k]-1$ and $\bbii_k[l] = \bbii[l]$ $(\forall l \ne k)$. Essentially, we look at one previous  index value in each dimension. This is sufficient because the table of $\gs_s$ and $\gs_e$ is built up progressively where later values accumulate the impact of all previous ones. 

In particular, we let  $\gs_s[\bbii] = \max\big\{\gs_s[\bbii], \min\{\gs_s[\bbii_k], 0 < k \le d-1\}\big\}$ ($\texttt{MinGS}(\cdot)$ at Line~\ref{sc:adjust}). This means that if the slice of \bbii does not overlap a key cell, we only need to start from the minimum $\gs_s$ of previous index combinations. This goes through $\spt(c)$ of candidate cell $c \in \cs_i$. 
Similarly, we let $\gs_e[\bbii] = \min\{\gs_e[\bbii_k]$, $0 < k \le d-1\}$ ($\texttt{MinGE}(\cdot)$ at Line~\ref{sc:adjust}). This is because $\gs_e$ of a previous index combination indicates a key cell found previously (Line~\ref{sc:appcse}). Cells at $[\bbii, cl_0]$ where $cl_0 > \min\{\gs_e[\bbii_k]$, $0 < k \le d-1\}$ will be at least partially dominated, and hence they can be pruned. 

\begin{figure*}[!htbp] 
  \centering
  \includegraphics[width=18cm]{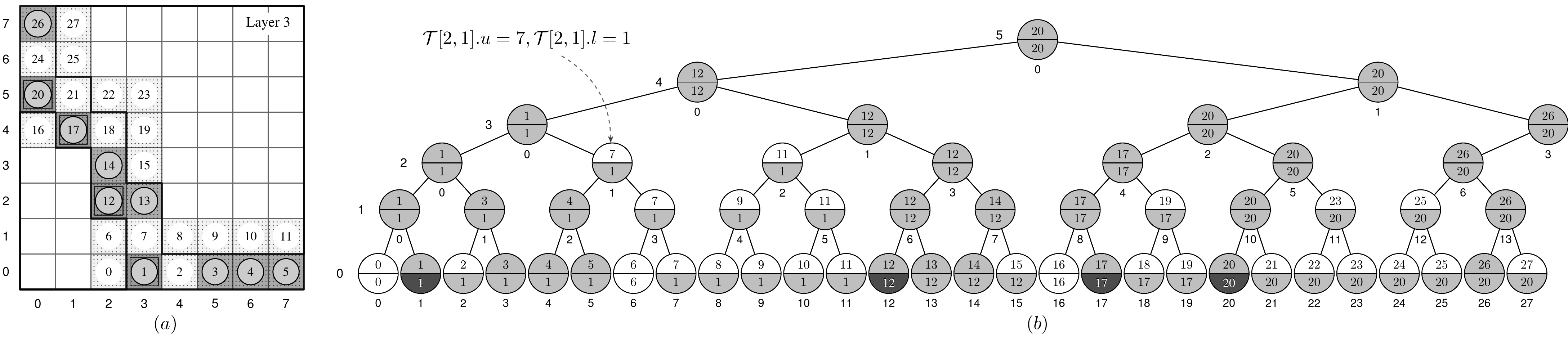}
  \caption{Example of parallel \texttt{ShrinkKeyCells}}
  \label{fig:egprl}
\end{figure*}

\textbf{Correctness.}
We next show the algorithm correctness.

\begin{lemma}\label{lm:mt}
  In Algorithm~\ref{alg:sc}, if $\ts_{i+1}[\bbii, cl_0] \in \ks_{i + 1}$, then $\gs_s[\bbii] \le cl_0 < \gs_e[\bbii]$.
\end{lemma}
\begin{proof}
  Suppose that there exists an $\bbii$ and a $cl_0$ such that $\ls_{i+1}[\bbii,\lb cl_0] \in \ks_{i+1}$ and $cl_0 < \gs_s[\bbii]$. Then, in $\ls_i$, there exists a candidate cell (note that a key cell is also a candidate cell) $\ls_i[\bbii^c, cl_0^c]$ such that $\ls_{i+1}[\bbii, cl_0] \in \spt(\ls_i[\bbii^c, cl_0^c])$. Let $\ls_{i+1}[\bbii^b,cl_0^b]$ be the cell with the smallest column indices 
  in $\spt(\ls_i[\bbii^k, cl_0^k])$ where $\ls_i[\bbii^k, cl_0^k]$ is any key cell that partially dominates $\ls_i[\bbii^c, cl_0^c]$. Then, $\bbii^b[j] \le \bbii[j], 0<j<d$ and $cl_0^b\le cl_0$. By function \texttt{MinGS}($\cdot$), $\gs_s[\bbii] \le \gs_s[\bbii^b] = cl_0^b\le cl_0$, contradicting  that $cl_0 < \gs_s[\bbii]$.

  Suppose there exists an $\bbii$ and a $cl_0$ such that $\ls_{i+1}[\bbii,\lb cl_0] \in \ks_{i+1}$ and $cl_0 > \gs_e[\bbii]$. By function $\texttt{MinGE}(\cdot)$, there exists a non-empty cell $\ls_{i+1}[\bbii^n, cl_0^n]$ such that $\bbii^n[j] \le \bbii[j], 0 < j < d$ and $cl_0^n = \gs_e[\bbii] < cl_0$. Then, $\ls_{i+1}[\bbii^n, cl_0^n]$ dominates $\ls_{i+1}[\bbii, cl_0]$, contradicting  that $\ls_{i+1}[\bbii, cl_0]$ is a key cell.
\end{proof}
    


\subsection{Parallel Key Cell Shrinking}
\label{sec:pave}
Next, we parallelize \texttt{ShrinkKeyCells}  with GPU.  
The algorithm takes $\cs_i$ and $\ks_i$ as the input. It generates $\spt(\cs_i)$ and then compares the cells, to find those not dominated by other cells and those not even partially dominated by other cells. This yields   
the candidate cells $\cs_{i+1}$ and the key cells $\ks_{i+1}$, respectively.
We parallelize the cell comparison with a tournament-style procedure. 

\textbf{Cell preparation.}
To generate cells in $\spt(\cs_i)$, we 
split each cell in $\cs_i$ into $2^d$ cells by an even split in each dimension. 
Unlike sequential \texttt{ShrinkKeyCells}, here, we do not prune the cells. 
We number the generated cells by their enumeration order (ascending). In Fig.~\ref{fig:egprl}a, the dotted cells denote the cells in $\spt(\cs_2)$ in Layer 3 (cf. Fig.~\ref{fig:egcells}). The number in cell denotes the \emph{cell number}. 

To compare the cells in $\spt(\cs_i)$ and identify those in $\cs_{i+1}$ and $\ks_{i+1}$, we construct an auxiliary binary tree $\mathcal{T}$. In this tree, each non-leaf node has two pointers $u$ and $l$ to point to the cells. Each leaf node has three pointers $o$, $u$, and $l$. 
We initialize both $o$ and $u$ of each leaf node (from left to right) to point to a cell in $\spt(\cs_i)$, in ascending order of the cell numbers. Fig.~\ref{fig:egprl}b shows such a tree for Fig.~\ref{fig:egprl}a. Every tree node (a circle) has two numbers. The upper (and lower) number represents the cell number of the cell pointed to by $u$ (and $l$). At start, only the upper half (pointer $u$) of the leaf nodes are labeled with cell numbers from 0 to 27 ($o$ points to the same cell as $u$ does and is not plotted). The rest of the nodes and pointers are computed later. The tree levels are numbered bottom-up, i.e., the leaf level is Level 0, which is denoted by $\mathcal{T}[0]$.

Note that  $\mathcal{T}$ is a complete binary tree 
with  $|\spt(\cs_i)| = 2^d |\cs_i|$ leaf nodes. 
By Corollary~\ref{thm:ratio}, the number of leaf nodes and the tree height $\mathcal{T}.h$ can be computed directly.  The tree is thus implemented as a one-dimensional array for fast parallel access. 

Algorithm~\ref{alg:prl2} summarizes  parallel  \texttt{ShrinkKeyCells}, 
where Line~\ref{sc:generate_all}  corresponds to the cell preparation steps above. 

\lcw{
  \begin{algorithm}[!htbp]
    \begin{small}
      \caption{ShrinkKeyCells (Parallel)}\label{alg:prl2}

      \SetAlgoNoLine
      \SetKwInOut{Input}{input}
      \SetKwInOut{Output}{output}
      \SetKwFunction{cost}{CostLine}
      \SetKwFunction{adjust}{Adjust}
      \SetKwFor{traverse}{traverse}{do}{end}
      \SetKwFor{forall}{for}{par-do}{end}
      \newcommand{\uu}[1]{\ensuremath{\mathcal{T}_u[#1]}}
      \newcommand{\dd}[1]{\ensuremath{\mathcal{T}_l[#1]}}

      \Input{\ Dataset $\ps$, current layer number $i$,
        $\cs_i$, $\ks_i$, $\ts_{\delta + 1}$}
      \Output{\ Candidate cells $\cs_{i+1}$, key cells  $\ks_{i+1}$}
      Assign $\spt(\cs_i)$ to pointers $o$ and $u$ for nodes in $\ttt[0]$ \label{sc:generate_all}\;
      \For{$m=0$ \KwTo $d-2$}{ \label{algl:rearrangeloop}
        Reorder $\ttt [0]$ by rotation-$m$\label{algl:rearrange}\;
       \If{$m \ne 0$} {
        \For{$j = 0$ \KwTo $|\spt(\cs_i)|$} {
        	 $\ttt[0, j].u \leftarrow \ttt[0, j].l$\;
       	}
       	}
        \For{$j=1$ \KwTo $\mathcal{T}.h$ }{\label{algl:r1}
          \forall{$k=0$ \KwTo $2^{\mathcal{T}.h-j} - 1$}{
            $\mathcal{T}[j,k].u  \leftarrow \mathtt{dom}_2(\mathcal{T}[j-1, 2k].u, \mathcal{T}[j-1, 2k+1].u)$\label{algl:comp}\;
          }
        }
        \For{$j=  \mathcal{T}.h$ \KwTo $0$}{\label{algl:r2}
          \forall{$k=0$ \KwTo $2^{\mathcal{T}.h-j} - 1$}{
            \If{$k=0$}{
              $\mathcal{T}[j,k].l \leftarrow \mathcal{T}[j, 0].u$\label{algl:case1}\; 
            }
            \ElseIf{$k$\emph{ is odd}}{
              $\mathcal{T}[j,k].l \leftarrow \mathcal{T}[j+1, (k-1)/2].l$
            }
            \Else{
              $\mathcal{T}[j,k].l \leftarrow \mathtt{dom}_2(\mathcal{T}[j+1, k/2-1].l, \mathcal{T}[j, k].u)$ \label{algl:dom}
            }
          }
        }
      } 
      $\cs_{i+1} \leftarrow \{\ttt[0, j].o | \ttt[0, j].l\pdom \ttt[0, j].o, 0 \le j < |\spt(\cs_i)|\}$\label{algl:candidate}\;
      $\ks_{i+1} \leftarrow \{\ttt[0, j].o | \ttt[0, j].l = \ttt[0, j].o, 0 \le j < |\spt(\cs_i)|\}$\label{algl:end}\;
      \Return $\cs_{i+1}$, $\ks_{i+1}$
    \end{small}
  \end{algorithm}
}

\textbf{Cell domination.} 
Next, we construct the upper levels of $\mathcal{T}$, during which cells in $\spt(\cs_i)$ 
are checked for domination. We first update pointer $u$ for the tree nodes bottom-up (Lines~\ref{algl:r1} to~\ref{algl:comp}). At tree level $j$ ($j$ starts at 1, i.e., parent nodes of the leaf nodes), let the $k$-th node be $\mathcal{T}[j, k]$. Its pointer $u$,  $\mathcal{T}[j, k].u$, will point to one of the two cells pointed to by the $u$ pointers of its two child nodes:
$$\mathcal{T}[j,k].u  = \mathtt{dom}_2(\mathcal{T}[j-1, 2k].u, \mathcal{T}[j-1, 2k+1].u) $$
Here, function $\mathtt{dom}_2(\cdot)$ checks for \emph{$2$-domination} (``$\sdom_2$'') between the cells pointed to by $\mathcal{T}[j-1, 2k].u$ and $\mathcal{T}[j-1, 2k+1].u$. 

\begin{definition}
  Given two cells $c_1$ and $c_2$,  if $c_1$ $k$-dominates $c_2$, denoted by 
  $c_1 \sdom_k c_2$, then $c_1 \sdom c_2$ and $\forall j \in [k, d-1], c_1[j] = c_2[j]$. Recall that 
  $\sdom$ denotes dominate or partially dominate.
\end{definition}

Intuitively, $k$-domination checks for domination (or partial domination, same below) in the lower $k$ dimensions. We use $\mathtt{dom}_2(\cdot)$ to check for two dimensions each time, and we rotate the dimensions such that all dimensions will be checked (Lines~\ref{algl:rearrangeloop} and~\ref{algl:rearrange}). In each rotation, dimension $k$ becomes dimension $k-1$ for $k > 0$, while dimension $0$ becomes the new dimension $d-1$ (nodes in $\mathcal{T}[0]$ is also reordered by the new column indices of the cells). A total of $d-1$ rotations are needed for $d$ dimensions.  
We only check for two dimensions each time to guarantee the algorithm correctness.

Function $\mathtt{dom}_2(\mathcal{T}[j-1, 2k].u, \mathcal{T}[j-1, 2k+1].u)$ returns $\mathcal{T}[j-1, 2k].u$ if it points to a cell that $2$-dominates the cell pointed to by $\mathcal{T}[j-1, 2k+1].u$. Otherwise, it returns $\mathcal{T}[j-1, 2k+1].u$. In Fig.~\ref{fig:egprl}, $\mathcal{T}[1, 0].u = \mathcal{T}[0, 1].u$, i.e., pointer $u$ of the left-most level-1 node should point to cell 1, because cell 0 is empty and it does not $2$-dominate cell 1.  The function is computed for every adjacent pair of nodes in parallel, as denoted by \textbf{par-do} in the algorithm. 

Computing the $u$ pointers bottom-up pushes the cells that dominate more cells to higher levels (e.g., the root node points to cell 20 which partially dominates 7 cells). 
Such cells may dominate more than just the cells in adjacent nodes. 
Next, we run a top-down procedure to check for domination between such cells and the cells in non-adjacent nodes, with the help of the $l$ pointers (Lines~\ref{algl:r2} to \ref{algl:dom}). At tree level $j$ ($j$ starts at $\ttt .h$), pointer $l$ of the $k$-th node is updated according to whether $k$ is $0$, odd, or even (detailed by Lemma~\ref{lm:U}).




After the $l$ pointers are updated, in the $m$-th rotation, 
for each node $n \in \mathcal{T}[0]$, $n.l$ points to a cell $c$ that $m+2$-dominates the cell pointed to by $n.o$, 
 while $c$ is not dominated by others (detailed in Lemma~\ref{lm:U}). 
 After $d-1$ rotations, for each node $n  \in \mathcal{T}[0]$, $n.l$ points to the cell that dominates or partially dominates the cell pointed to by $n.o$ (if there exists such a cell). 
The cell pointed to by $n.o$ is a candidate cell if $n.l \ne n.o$ (Line~\ref{algl:candidate}), 
and a key cell otherwise (Line~\ref{algl:end}, e.g., cells $1$, $12$, $17$, and $20$ 
in Fig.~\ref{fig:egprl}).

\textbf{Correctness.} Next, we prove the algorithm correctness.  
We use $o$, $l$, and $u$ to refer to the cells pointed to by them in the discussion. 
Our proof is built on the following four lemmas. 

\begin{lemma}\label{lm:sdom}
  Given two cells $c'$ and $c$, if $c' \sdom c$, then $c' \sdom \mu_i(c', c), \forall 0 < i < d$, where $\mu_i(c', c)$ is a cell where $\mu_i(c', c)[j] = c'[j], \forall 0 \le j < i$ and $\mu_i(c', c)[j] = c[j], \forall i \le j < d$.
\end{lemma}
\begin{proof}
Straightforward based on Definition~\ref{df:cd}.
\end{proof}

We define sets $O(n)$ and $L(n)$ for tree node $n$: 
$O(n)$ is the set of $o$ cells in the leaf nodes of the subtree rooted at $n$; $L(n)$ 
includes $O(n)$ and all $o$ cells in the leaf nodes preceding the subtree rooted at $n$. 
$$O(\tul{j,k})=\bigcup_{i=2^j k}^{2^{j} (k+1)-1}\{\too{0,i}\};\, L(\tul{j,k})=\bigcup_{i=0}^{2^{j} (k+1)-1}\{\too{0,i}\}$$
For example, $O(\mathcal{T}[2,1]) = \{c_4, c_5, c_6, c_7\}$, $L(\mathcal{T}[2,1]) = \{c_0, \ldots, c_7\}$.

Given a list of cells $C$, we define $\ds(C,k)$ as the cell that (i) belongs to $C$, and (ii) $k$-dominates the last cell of $C$, 
and (iii) is not $k$-dominated by any other cell in $C$. Here, the last cell is defined by sorting cells in $C$ by their column indices in ascending order with the current rotation of the dimensions. For example, if $C = \{c_0, c_1, c_2, c_3\}$ in Fig.~\ref{fig:egprl}, the last cell is $c_3$ and $\ds(C, 2)$ is $c_1$.

These definitions help show a property of the $o$ cells for $m=0$.
\begin{lemma}\label{lm:u}
   In Algorithm~\ref{alg:prl2}, at the end of the iteration for $m=0$, each node $\tul{j,k}$ satisfies $\tu{j,k}=\ds(O(\tul{j,k}), 2)$.
\end{lemma}
\begin{proof}
  For $j=0$, each node $\tul{0,k}$ is in its own subtree, i.e. $O(\tul{0,k}) = \{\too{0,k}\}$. Therefore, $\tu{0,k} = \ds(O(\too{0,k}), 2)$ by definition (recall that $\tu{0,k} = \too{0,k}$ when $m=0$).

  When nodes $\tul{j,\cdot}$ satisfy the lemma, we show that nodes $\tul{j+1,\cdot}$ also satisfy the lemma. At Line~\ref{algl:comp} of the algorithm, we have $\tu{j+1, k} = \mathtt{dom}_2(\tu{j, 2k}, \tu{j, 2k + 1})$. Since $\tul{j+1, k}$ is the parent of $\tul{j, 2k}$ and $\tul{j, 2k + 1}$, $O(\tul{j+1, k}) = O(\tul{j, 2k})\cup O(\tul{j, 2k+1})$. Function $\mathtt{dom}_2(\tu{j, 2k},\lb \tu{j, 2k + 1})$ yields either $\tu{j, 2k}$ or $\tu{j, 2k+1}$. Thus, $\tu{j+1, k}$ satisfies condition (i) of $\beta(\cdot)$. We have two cases for the other conditions:

  Case 1: $\mathtt{dom}_2(\tu{j, 2k}, \tu{j, 2k + 1}) = \tu{j, 2k + 1}$, i.e. $\tu{j, 2k} \not\sdom_2 \tu{j, 2k + 1}$. We have $\tu{j, 2k}[0] > \tu{j, 2k + 1}[0]$ and $\tu{j, 2k}[1] < \tu{j, 2k + 1}[1]$. Condition (ii) holds as the last node of $O(\tul{j, 2k+1})$ is also the last node of $O(\tul{j+1, k})$. We prove condition (iii) by contradiction. Suppose there exists another cell $c_p \in O(\tul{j+1, k})$ that 2-dominates $\tu{j, 2k+1}$. Then, $c_p\in O(\tul{j, 2k})$. If $c_p = \tu{j, 2k}$, then $\tu{j,2k}\sdom_2 \tu{j, 2k +1}$, and we have a contradiction. If $c_p$ is before $\tu{j, 2k}$, then $\tu{j, 2k}[0] < c_p[0]$ and $\tu{j, 2k}[1] > c_p[1]$ (otherwise $c_p \sdom_2 \tu{j, 2k}$). Since the cells are ordered by column indices, $\tu{j, 2k}[1] \le \tu{j, 2k+1}[1]$. Since $c_p \sdom_2 \tu{j,2k+1}$, $c_p[0] \le \tu{j,2k+1}[0]$, $\tu{j, 2k}\sdom_2 \tu{j, 2k+1}$, and we have a contradiction. If $c_p$ is after $\tu{j, 2k}$, then $c_p$ will 2-dominate the last cell of $O(\tul{j,2k})$, contradicting with $c_p \ne \tu{j, 2k}$.  

  Case 2: $\mathtt{dom}_2(\tu{j, 2k}, \tu{j, 2k + 1}) = \tu{j, 2k}$, that is, $\tu{j, 2k} \sdom_2 \tu{j, 2k + 1}$. Condition (ii) holds as $\tu{j, 2k} \sdom_2 \tu{j, 2k +1} \sdom_2 c_l$. Here, $c_l$ is the last cell of $O(\tul{j, 2k+1})$. Condition~(iii) holds because: $\tu{j,2k}$ is not 2-dominated by any other cell in $O(\tul{j, 2k})$; and $\tu{j, 2k}$ cannot be 2-dominated by any cell in $O(\tul{j, 2k+1})$. The later is because cells in $O(\tul{j, 2k+1})$ are positioned after $\tu{j, 2k}$. When ordered by column indices, cell $c_a$ can dominate cell $c_b$ only if $c_a$ is before $c_b$.  
\end{proof}

We also show a property of the $l$ cells of the leaf nodes for $m=0$.

\begin{lemma}\label{lm:l}
  In Algorithm~\ref{alg:prl2}, at the end of the iteration for $m=0$, each node $\tul{j,k}$ satisfies $\tl{j,k}=\ds(L(\tul{j,k}), 2)$.
\end{lemma}
\begin{proof}
  At Line~\ref{algl:r2}, for $j=\ttt.h$, there is only one node $\tul{\ttt.h, 0}$, and $O(\tul{\ttt.h, 0}) = L(\tul{\ttt.h, 0})$. Let $\tl{\ttt.h, k} = \tu{\ttt.h, k}$ for the only value $k = 0$ at Line\ref{algl:case1}. Then, $\tl{j, k} = L(\tul{j, k})$.

  When nodes $\tl{j,\cdot}$ satisfy the lemma, we show that nodes  $\tl{j-1,\cdot}$ also satisfy the lemma. There are three cases of $k$:

  (1) $k=0$: $O(\tul{j-1, k}) = L(\tul{j-1, k})$, we set $\tl{j-1, k} = \tu{j-1, k}$. Since $\tu{j-1, k}$ is $\ds(O(\tul{j-1, k}), 2)$, $\tl{j-1, k}$ is $\ds(L(\tul{j-1, k}), 2)$.

  (2) $k$ being odd: In this case, $\tul{j-1, k}$ is the right child of $\tul{j, (k-1)/2}$. Then,  $L(\tul{j-1, k}) = L(\tul{j, (k-1)/2})$. Thus, we can simply set $\tl{j-1, k} = \tl{j, (k-1)/2}$.

  (3) $k$ being even: In this case, $L(\tul{j-1, k}) = L(\tul{j, k/2-1}) \cup O(\tul{j-1, k})$. We set $\tl{j-1,k} = \mathtt{dom}_2(\tl{j, k/2-1}, \tu{j-1,k})$. This satisfies the three conditions in a way similar to the two cases in Lemma~\ref{lm:u}. We omit the detail reasoning. 
\end{proof}

We generalize the results to later iterations for $m \ge 0$.

\begin{lemma}\label{lm:U}
  In Algorithm~\ref{alg:prl2}, after the $m$-th iteration (Lines~\ref{algl:rearrange} to \ref{algl:dom}), for each leaf node $n$, $n.l = \ds(L(n), m+2)$.
\end{lemma}
\begin{proof}
  The lemma holds when $m=0$. Suppose that the lemma holds when $m=\alpha$. We prove that it also holds when $m=\alpha+1$.

  Let $n(c)$ be the leaf node where $n.o = c$. After the $\alpha$-th iteration, for any cell $c$ that is $(\alpha+3)$-dominated, $n(c).l$ is the cell that $(\alpha+2)$-dominates $c$ and is not dominated by any other cell. Further, $n(c).l[\alpha+2] = c[\alpha+2]$. Then, we know that, if $n(c).l$ is not dominated by other cells, then $n(c).l$ $(\alpha+3)$-dominates $c$. If $n(c).l$ is $(\alpha+3)$-dominated by another cell $c'$, i.e., $c'\sdom_{\alpha+3} n(c).l$, after the $\alpha$-th iteration, $n(c_\mu).l$ is now $c'$, where $c_\mu=\mu_{\alpha+2}(c', n(c).l)$. This can be proven as $c'$ satisfies the three conditions. Under rotation $\alpha+1$, $c_\mu$ is positioned before $c$. Then, after the $(\alpha+1)$-th iteration, $n(c).l$ will be replaced at  $n(c_\mu).l$, which is $c'$. Thus, $n.l = \ds(L(n), \alpha+3)$.
\end{proof}

Now we can show the algorithm correctness with Theorem~\ref{thm:prl2d}.

\begin{theorem}\label{thm:prl2d}
  At  the end of Algorithm~\ref{alg:prl2}, $\too{0, k}$ is a candidate cell if and only if $\tl{0,k}\pdom \too{0,k}$, and it is a key cell if and only if $\tl{0,k}=\too{0,k}$.
\end{theorem}

\begin{proof}
  By Lemma~\ref{lm:U}, at the end of the algorithm, for each leaf node $n$, $n.l$ is $\beta(L(n), d)$, which is the cell that dominates or partially dominates cell $n.o$ or is $n.o$. When a cell is partially dominated but not dominated by others, it is a candidate cell. When a cell is neither dominated nor partially dominated, it is a key cell.
\end{proof}

\section{Cost Analysis}
\label{sec:ca}

For SkyCell (Algorithm~\ref{alg:sq}), sorting $n$ points takes 
$\bo(n\log n)$ time.
Constructing grid layers $\ls_\rho$ to $\ls_{0}$ takes $\bo(2^{\rho\cdot d})$ time, where $2^{\rho\cdot d}$ is the number of cells in $\ls_\rho$.
When the points are distributed evenly, $\rho$ is at most $(\log n)/d$, with one point per cell in $\ls_\rho$.

Then,  \texttt{ShrinkKeyCells} is run from Layer 0 to Layer $\rho - 1$. 
In sequential \texttt{ShrinkKeyCells} (Algorithm~\ref{alg:sc}), we go through a subset of candidate cells in Layer $i$ to generate key cells in Layer $i+1$. For each cell, 
an adjustment procedure is run to update the column index range to be enumerated, which takes $\bo(\log d)$ time. 
By Corollary~\ref{thm:ratio}, in Layer $i$, the number of candidate cells is $\sum_{j=0}^{d-1}(2^i-1)^j 2^{i(d-1-j)}$. Thus, the time complexity of Algorithm~\ref{alg:sc} is:
  \begin{equation}
    \begin{array}{l}
      \bo\left(\log d\sum_{i=0}^{\rho-1}\sum_{j=0}^{d-1}(2^i-1)^j  2^{i(d-1-j)}\right)\\
      =\bo\left(\log d\sum_{i=0}^{\rho-1}2^{i\cdot d} \right) = \bo(2^{\rho \cdot d} \log d)
    \end{array}
  \end{equation}

In parallel \texttt{ShrinkCandidates} (Algorithm~\ref{alg:prl2}), trees $\mathcal{T}_u$ and $\mathcal{T}_l$ are updated $d-1$ times, each taking a logarithmic time  to the number of candidate cells. Therefore, the algorithm time complexity is: 
  \begin{equation}
      \begin{array}{l}
    \bo\left((d-1)\sum_{i=0}^{\rho-1}\log \sum_{j=0}^{d-1} (2^i-1)^j 2^{i (d-1-j)}\right)\\
    =\bo\left( \rho \cdot d \cdot \log 2^{\rho \cdot d} \right) = \bo(\rho^2 \cdot d^2)
        \end{array}
  \end{equation}

  Since there are only a few points (mostly skyline points) in each candidate cell in $\cs_\rho$, and the cells can be processed in parallel, \texttt{RefineSkyline}  has roughly a quadratic time 
  to the number of points in each cell. Each cell is expected to contain $n/ 2^{\rho\cdot d}$ points, and \texttt{RefineSkyline} takes $\bo(n^2/2^{2\rho\cdot d})$ time. 

Overall, when $\rho=(\log n)/d$ (the maximum value given $n$ uniformly distributed points), the time complexity of our sequential and parallel algorithms are $\bo\left( n \log d \right)$ and $\bo\left(\log^2 n  \right)$, respectively.

Our grid take $\bo(2^{\rho \cdot d})$ space. Our sequential method further stores the column index bounds for $\bo(2^{\rho (d-1)})$ cells.
Parallel \texttt{ShrinkKeyCells} stores $\sum_{i=1}^{\log |\cs_\rho|}{|\cs_\rho|/2^i}$ cells for the tree structure, where $|\cs_\rho|=\sum_{j=0}^{d-1} (2^\rho-1)^j 2^{\rho (d-1-j)}$,  yielding an $\bo(\rho \cdot d \cdot 2^{\rho \cdot d})$ cost. When $\rho=(\log n)/d$, the space complexity of our sequential and parallel algorithms are $\bo(n)$ and $\bo(n\log n)$, respectively. 

For comparison, the state-of-the-art sequential skyline algorithm~\cite{liu2018skyline} takes $\bo(n^4 \log n)$ time and $\bo(n^{2d+1})$ space. 

\section{Experimental Evaluation}
\label{sec:ee}
We compare with three state-of-the-art algorithms, \textbf{Skyline Diagram}~\cite{liu2018skyline}. \textbf{Hybrid}~\cite{chester2015scalable} and \textbf{SkyAlign}~\cite{bogh2015work}. Specifically, we compare the sequential version of our algorithm \textbf{SkyCell} with Skyline Diagram which is a sequential algorithm on CPU. We compare the parallel version of \textbf{SkyCell} with Hybrid and SkyAlign which are parallel skyline algorithms on GPU. 

\subsection{Settings}
\label{sec:st}
We implement all algorithms with C++ and CUDA 10.0 (code available on GitHub~\cite{github}).   
We use a 64-bit machine with 32 GB memory, a 2.1 GHz Intel Xeon Silver 4110 CPU (8 cores), and an Nvidia Quadro RTX6000 GPU with 4,608 cores and 24 GB memory.


\textbf{Datasets.} 
We obtain 3.2 billion data points ($d=2$) from OpenStreetMap~\cite{osm} to form a real dataset, denoted by ``\textbf{OSM}''. We create subsets by random sampling for experiments on varying dataset cardinality. We further synthesize real datasets by using randomly sampled  coordinates from the first two dimensions as coordinates in the higher dimensions. 
We also generate synthetic data using a commonly used dataset generator~\cite{borzsony2001skyline} following previous studies~\cite{bogh2015work, chester2015scalable, liu2018skyline}. The datasets generated include \textbf{Independent}, \textbf{Anti-correlated}, and \textbf{Correlated}, where the coordinates of a point in different dimensions are independent, anti-correlated, and correlated, respectively. We vary the data dimensionality $d\in [1, 10]$, and dataset cardinality $n\in\{1, 2, \ldots,32\}\times 10^8$ for parallel implementation and $n\in\{1, 2, \ldots,10\}\times 10^6$ for sequential implementation. By default, we set $d=4$ and $n=4\times 10^8$ for the parallel algorithms; we set $d=2$ and $n=2\times 10^6$ for the sequential algorithms. We measure and report the algorithm running times.


\subsection{Results}
\label{sec:rs}

We first report the impact of the partition ratio $\rho$ in Section~\ref{sec:op}, to guide the choice of its value for the later experiments. 
Then, we report the performance of the parallel and the sequential algorithms in Sections~\ref{sec:csa} and~\ref{sec:csd}, respectively. 

\def\eheight{2.75}
\def\lheight{3.3}
\def\wdh{.20}
\def\wdhd{.36}
\def\wdh{.18}

\begin{figure}[!hbpt]
  \centering 
  \subfloat[\scriptsize The ratio of candidate cells vs. $\rho$]{\includegraphics[width=\wdh\textheight,height=70pt]{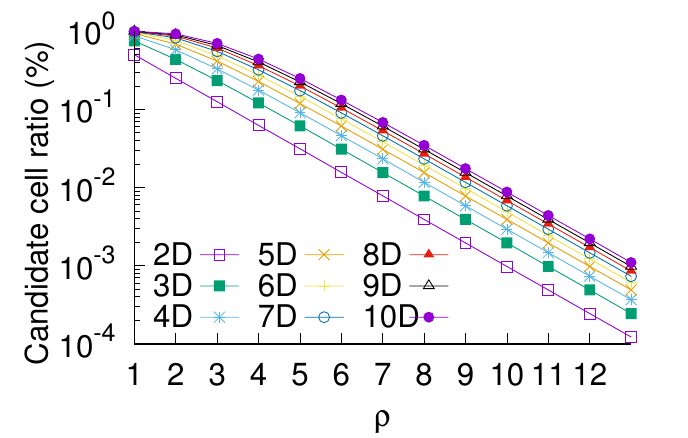}}\hspace{5pt}
  \subfloat[\scriptsize Time for \texttt{ShrinkKeyCells}]{\includegraphics[width=\wdh\textheight,height=70pt]{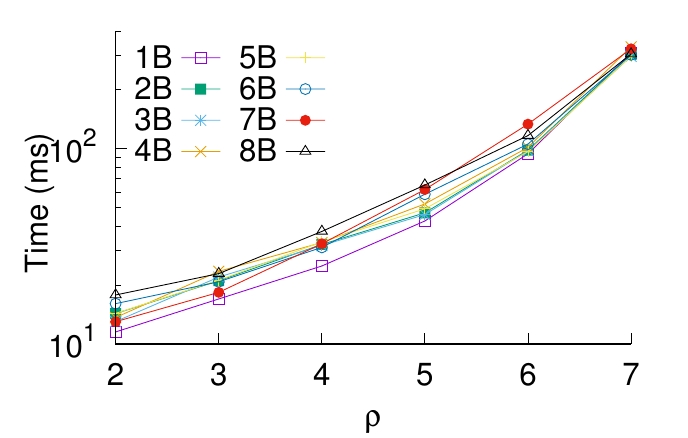}}\\
  \subfloat[\scriptsize Time for \texttt{RefineSkyline}]{\includegraphics[width=\wdh\textheight,height=70pt]{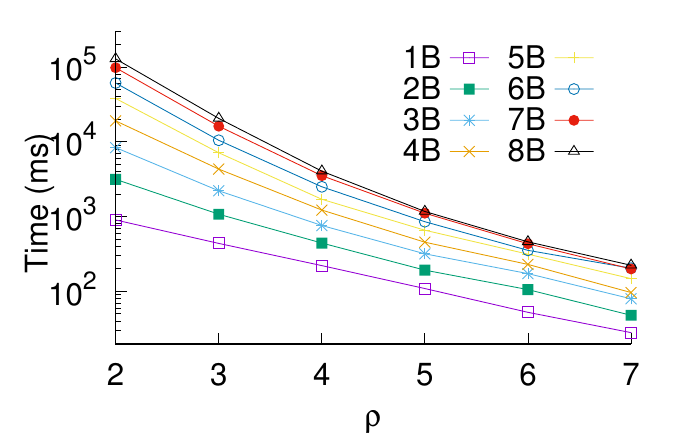}}\hspace{5pt}
  \subfloat[\scriptsize Overall running time]{\includegraphics[width=\wdh\textheight,height=70pt]{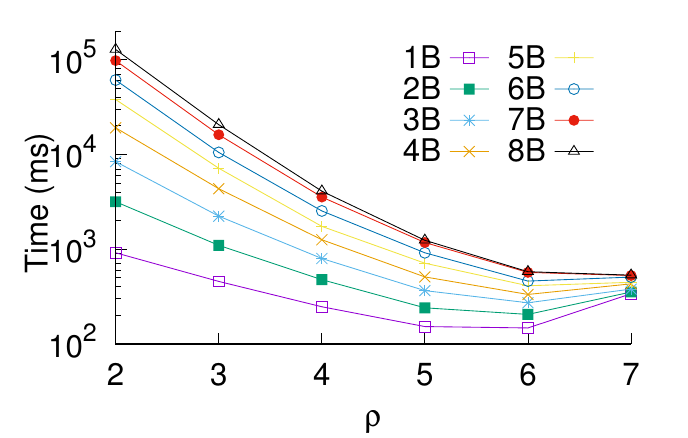}}
  \caption{The impact of partition ratio $\rho$} 
  \label{fig:rho_src} 
\end{figure}

\subsubsection{Impact of Partition Ratio} 
\label{sec:op}
Fig.~\ref{fig:rho_src}a shows the ratio of Layer $\rho$ (in our multi-layer grid) being covered by candidate cells, as  computed by Corollary~\ref{thm:ratio}, for $\rho \in [1, 12]$ and $d \in [2, 10]$. Note that this ratio depends only on the layer number and $d$, and is independent from the dataset cardinality and distribution. We can see that the ratio of the space covered by the candidate cells decreases exponentially (note the logarithmic scale) with the increase of $\rho$. When $d=2$, the candidate cells cover less than 1\% of the space at $\rho = 7$, and this ratio further drops to 0.01\% at $\rho = 12$. When $d=10$, we still just need $\rho = 10$ so that the candidate cells only cover 1\% of the data space. These results verify that our SkyCell algorithm can quickly prune a large portion of the data space (and hence the data points) from consideration with grids of only a few layers.


We further show in Fig.~\ref{fig:rho_src} the overall algorithm running time, the time for key cell shrinking, and the time for refinement (skyline point computation), as $\rho$ varies from $2$ to $7$ over Independent data  with $1$ billion (``1B'') to $8$ billion (``8B'') points (for parallel SkyCell and $d= 4$). As $\rho$ increases, the time for key cell shrinking increases (Fig.~\ref{fig:rho_src}b), while that for skyline point computation decreases (Fig.~\ref{fig:rho_src}c), which are both expected. Their combined effect (Fig.~\ref{fig:rho_src}d), is an optimal overall running time at $\rho = 6$. Also, as $n$ increases, grids with a larger resolution (i.e., larger $\rho$)  help prune more points from further checking. Thus, the curve of 8B drops faster than that of 1B with the increase of $\rho$. 
The algorithm performance on other settings shows a similar pattern. We thus use $\rho = 6$ as the default value.

\subsubsection{Performance of Parallel SkyCell}
\label{sec:csa}

\begin{figure}[!hbt]
  \centering 
  \subfloat[\scriptsize Independent]{\includegraphics[width=\wdh\textheight,height=70pt]{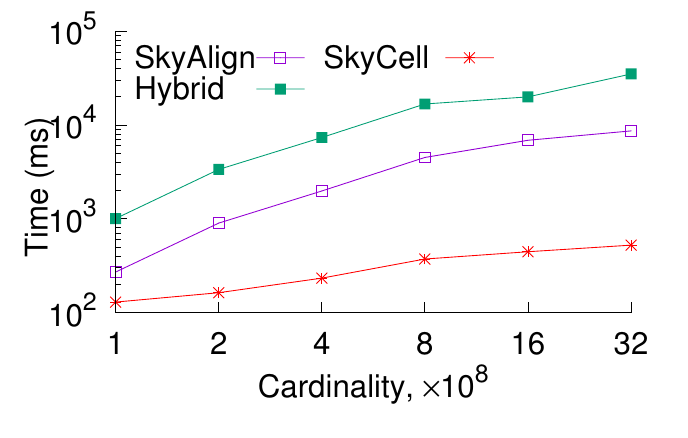}}\hspace{5pt}
  \subfloat[\scriptsize Anti-correlated]{\includegraphics[width=\wdh\textheight,height=70pt]{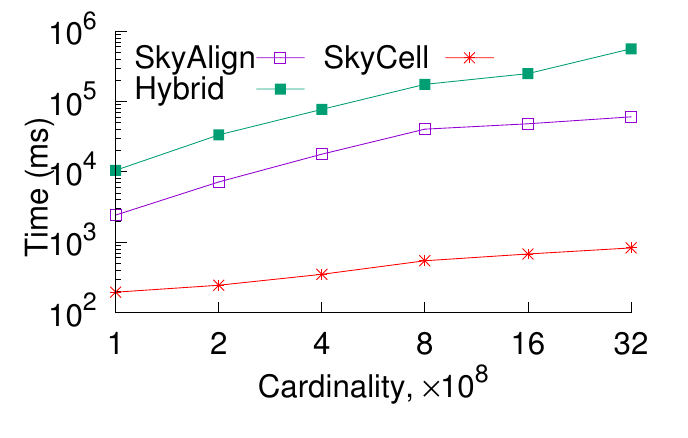}}\\
  \subfloat[\scriptsize Correlated]{\includegraphics[width=\wdh\textheight,height=70pt]{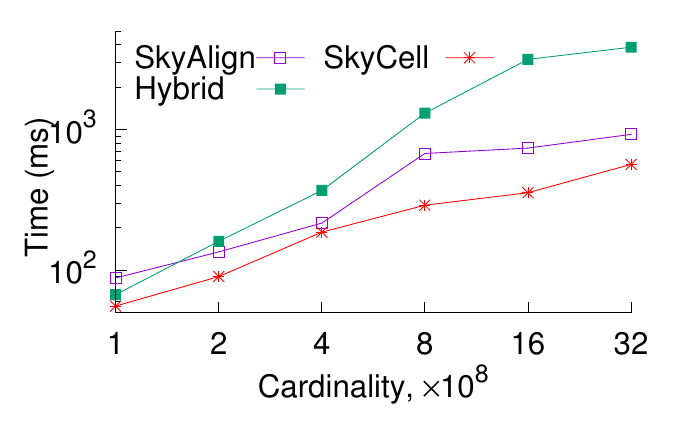}}\hspace{5pt}
  \subfloat[\scriptsize OSM]{\includegraphics[width=\wdh\textheight,height=70pt]{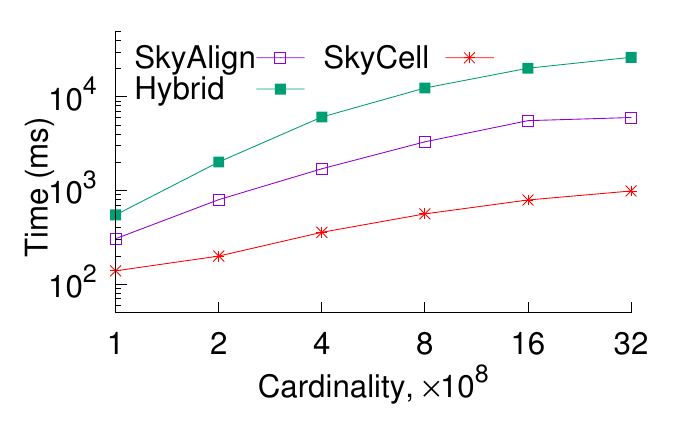}}
\caption{Performance of parallel SkyCell vs. $n$}
  \label{fig:sa} 
\end{figure}

\begin{figure}[!hbt]
  \centering 
  \subfloat[\scriptsize Independent]{\includegraphics[width=\wdh\textheight,height=70pt]{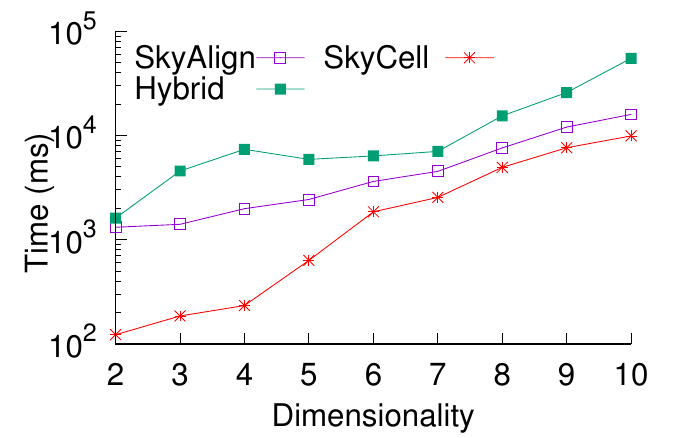}}\hspace{5pt}
  \subfloat[\scriptsize Anti-Correlated]{\includegraphics[width=\wdh\textheight,height=70pt]{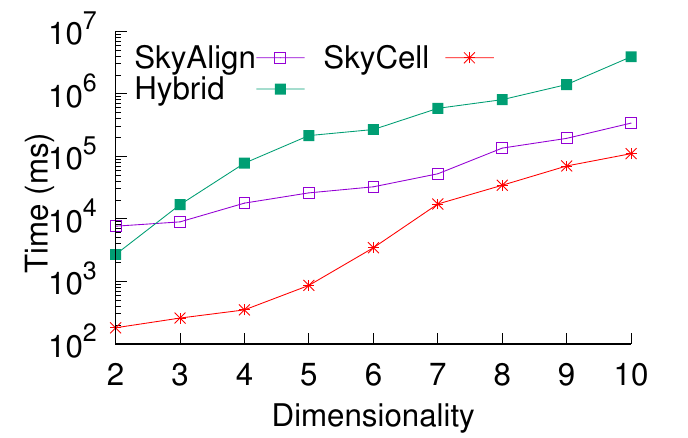}} \\
  \subfloat[\scriptsize Correlated]{\includegraphics[width=\wdh\textheight,height=70pt]{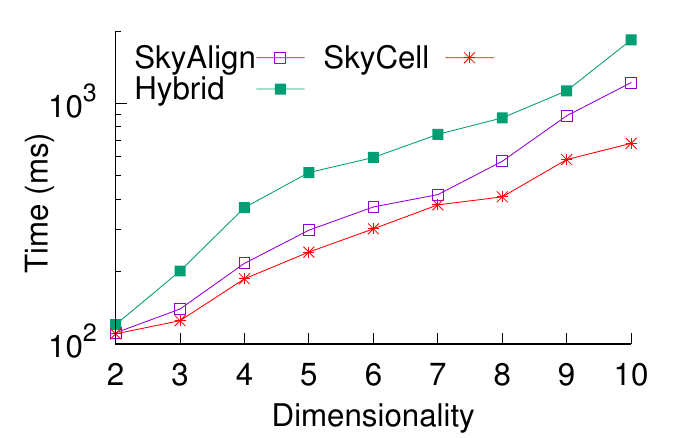}}\hspace{5pt}
  \subfloat[\scriptsize OSM]{\includegraphics[width=\wdh\textheight,height=70pt]{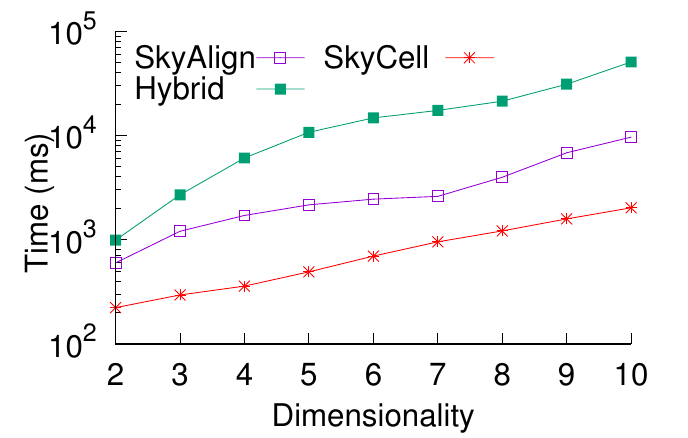}}
\caption{Performance of parallel SkyCell vs. $d$}
  \label{fig:sa_d} 
\end{figure}

\begin{figure}[!hbt]
  \centering 
    \subfloat{\includegraphics[width=\wdh\textheight,height=70pt]{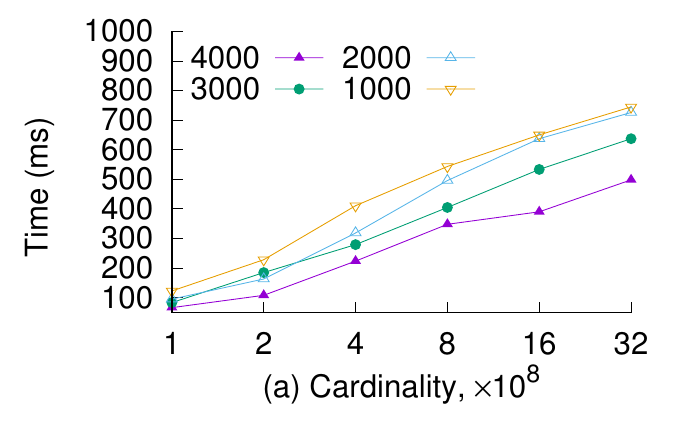}}\hspace{5pt}
  \subfloat{\includegraphics[width=\wdh\textheight,height=70pt]{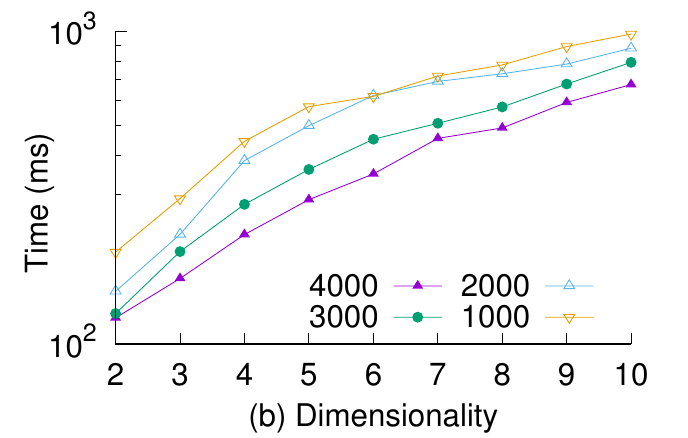}}
\caption{Performance of parallel SkyCell vs. \# GPU threads}
  \label{fig:pt} 
\end{figure}

We show the results for of parallel algorithms in Figs.~\ref{fig:sa} to~\ref{fig:pt}. 

\emph{Impact of dataset cardinality $n$.} We see that the algorithm running times increase with $n$ (Figs.~\ref{fig:sa}). Our SkyCell algorithm outperforms SkyAlign and Hybrid 
consistently on both synthetic and real data. Its running times are more stable (between 100 and 1,000 ms) across datasets of different distributions. This is because its cell-based pruning strategy of SkyCell is more robust against the data distribution. 
On Independent and Anti-correlated data, in general, SkyCell outperforms SkyAlign and Hybrid by one and two orders of magnitude (up to 60 and 700 times), respectively. On Correlated data, SkyAlign and Hybrid become closer to (but still worse than) SkyCell.  There are few skyline points on such data (e.g., 4,203 points among $32 \times 10^8$ data points, and many data points can be pruned by a skyline point, which benefit the point-based algorithms SkyAlign and Hybrid. Even in this extreme case, SkyCell runs the fastest. It computes the skyline points from $32\times 10^8$ points in just about 0.6  seconds.  On OSM, SkyCell outperforms SkyAlign and Hybrid by 6 and 27 times at $n = 32 \times 10^8$, where it finishes in under a second. 
These confirm the scalability of our algorithm.

\emph{Impact of data dimensionality $d$.} In Fig.~\ref{fig:sa_d}, we vary $d$ from 2 to 10. The algorithm running times increase with $d$ in general, while Hybrid has a fluctuation which is also observed in its original proposal~\cite{chester2015scalable}. SkyCell again outperforms the competitors on all datasets consistently. When $d=10$, comparing with Hybrid, SkyCell reduces the running time by 82\%, 97\%, 62\%, and 96\% on the  Independent, Anti-correlated, Correlated, and OSM data, respectively. When comparing with SkyAlign, these numbers become 38\%, 67\%, 44\%, and 79\%, respectively. 
We observe that, on  Independent and Anti-correlated data, SkyAlign and Hybrid become closer to SkyCell for $d \ge 6$. This is because SkyCell becomes less optimal with its default $\rho$ value under these settings.  Even in this less optimal case,  SkyCell still outperforms SkyAlign and Hybrid, confirming its robustness against the choice of $\rho$ value. 


\emph{Impact of number of threads.} 
In Fig.~\ref{fig:pt}, we test the capability of SkyCell to exploit the parallel power of GPU by running the algorithm on datasets of different cardinality and dimensionality while varying  the number of threads used on the GPU from 1,000 to 4,000. We see that, given fixed dataset cardinality and dimensionality, 
the running time of SkyCell decreases almost linearly with the increase in the number of threads. This confirms  the capability of SkyCell to take full advantage of the parallel processing power of GPU. 


\subsubsection{Performance of Sequential SkyCell}
\label{sec:csd} 

We compare our sequential SkyCell algorithm with the \textbf{Scan} and \textbf{Sweep} algorithms using the Skyline Diagram technique~\cite{liu2018skyline}. Following Skyline Diagram, we set up the skyline queries dynamically (i.e., the quadrant skyline query), where a query point with random coordinates is used as the new origin of the data space. Only data points on the top-right quadrant are considered for skyline computation. 

We generate $n_q = 10,000$ queries on each dataset and report the average algorithm response time. For Scan and Sweep, since they require 
pre-computation, we amortize the pre-computation time into the algorithm response time $t$, 
i.e.,  $t = (t_p + t_q) / n_q$, where $t_p$ is the pre-computation time and $t_q$ is the query time. 

\begin{figure}[!hbt]
  \centering 
  \subfloat[\scriptsize Independent]{\includegraphics[width=\wdh\textheight,height=70pt]{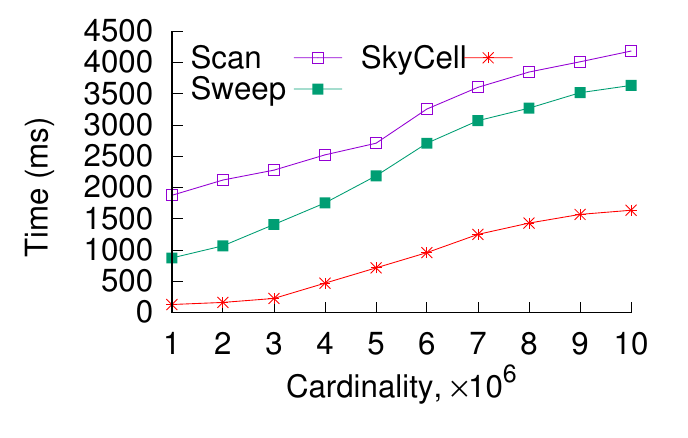}}\hspace{5pt}
  \subfloat[\scriptsize Anti-correlated]{\includegraphics[width=\wdh\textheight,height=70pt]{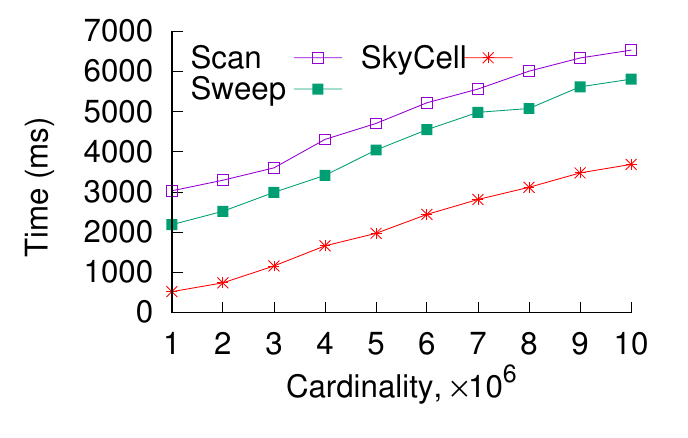}}\hspace{5pt}\\
  \subfloat[\scriptsize Correlated]{\includegraphics[width=\wdh\textheight,height=70pt]{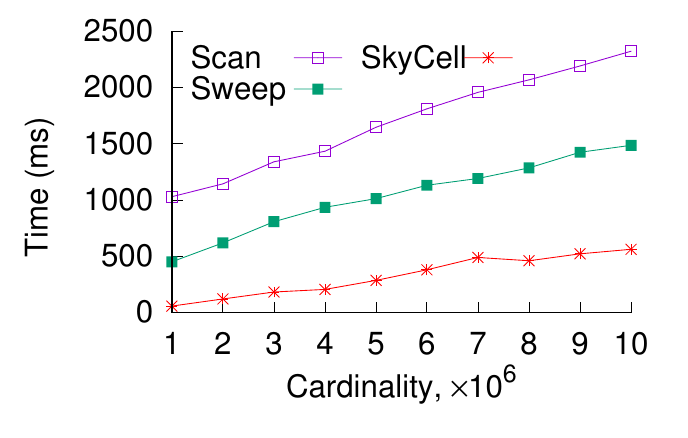}}\hspace{5pt}
    \subfloat[\scriptsize OSM]{\includegraphics[width=\wdh\textheight,height=70pt]{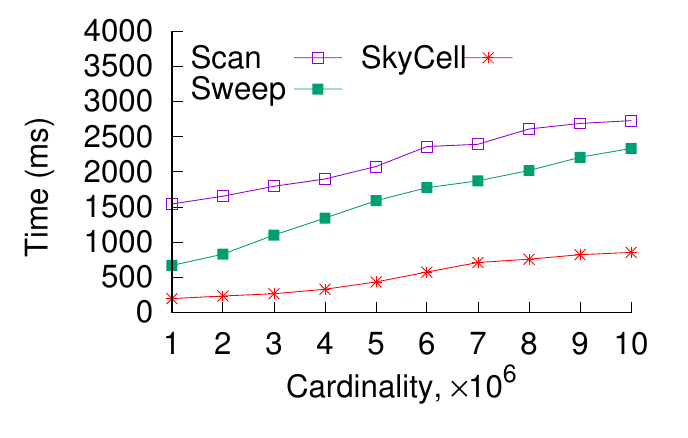}}\hspace{5pt}
  \caption{Performance of sequential SkyCell vs. $n$} 
  \label{fig:sd} 
\end{figure}

\emph{Impact of dataset cardinality $n$.} 
Fig.~\ref{fig:sd} presents the algorithm running times as $n$ varies from $1\times 10^6$ to $10 \times 10^6$. We see that SkyCell outperforms both Scan and Sweep consistently, and the advantage is up to 4 and 8 times, respectively. To be fair, this is because the pre-computation times of Scan and Sweep have been amortized into the running times. We argue that a skyline algorithm that requires a heavy pre-computation suffers in its applicability, because real datasets are often dynamic where updates (e.g., data insertions and deletions) may invalidate the pre-computed results. Our SkyCell algorithm does not have such a limitation. It applies to both static and dynamic scenarios with a high efficiency, e.g., computing skyline points from 10 million real data points in less than 1 second as shown in Fig.~\ref{fig:sd}d.

\begin{figure}[!hbt]
  \centering 
\subfloat[\scriptsize Independent]{\includegraphics[width=\wdh\textheight,height=70pt]{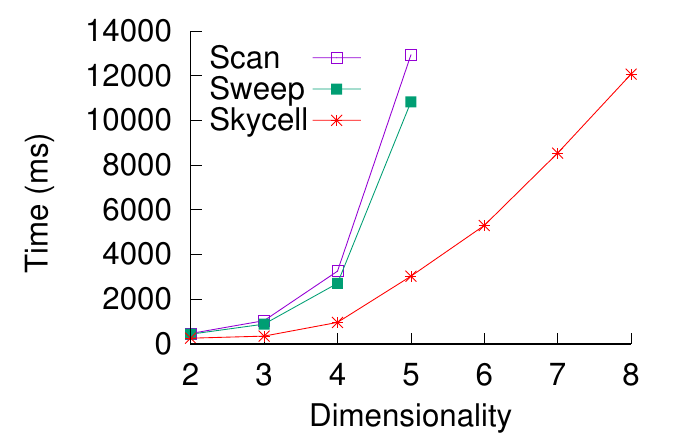}}\hspace{5pt}
  \subfloat[\scriptsize Anti-correlated]{\includegraphics[width=\wdh\textheight,height=70pt]{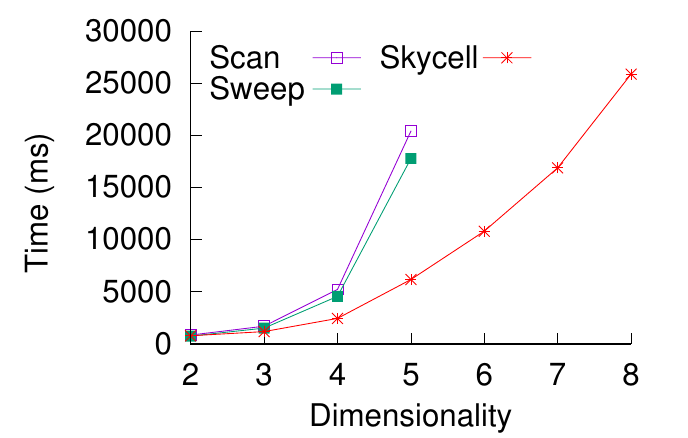}}\\
  \subfloat[\scriptsize Correlated]{\includegraphics[width=\wdh\textheight,height=70pt]{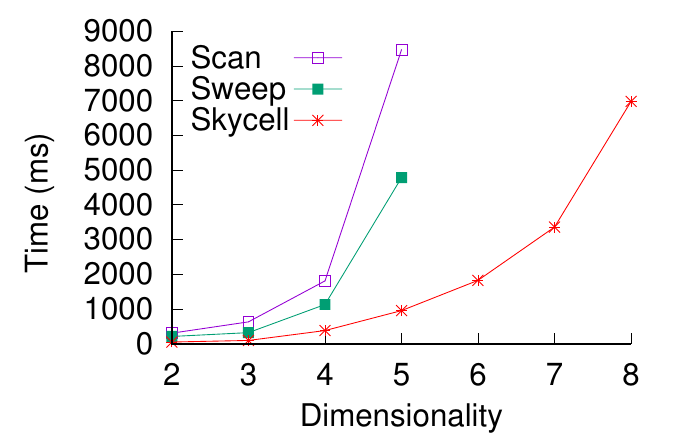}}\hspace{5pt}
  \subfloat[\scriptsize OSM]{\includegraphics[width=\wdh\textheight,height=70pt]{exp/dim_linear_cor.pdf}}
  \caption{Performance of sequential SkyCell vs. $d$} 
  \label{fig:sd_d} 
\end{figure}

\emph{Impact of data dimensionality $d$.} 
In Fig.~\ref{fig:sd_d}, we vary $d$ up to 8 (instead of 10 where the algorithms take too long to run). SkyCell again outperforms the competitors, and the advantage grows with $d$. As discussed earlier, when $d$ increases, although the key cell shrinking takes more time, the refinement stage may take less time (as the candidate cells occupy a smaller  space). In contrast, Scan and Sweep need to process exponentially more cells with the increase of $d$, which brings rapidly increasing running times. Moreover, Scan and Sweep needs to store much pre-computation data. Their pre-computation cannot finish in 4 hours on our hardware  for $d > 5$, and hence no results have been reported for them in these cases.  

\section{Conclusions}
\label{sec:conc}
We studied skyline queries and proposed a grid structure 
that enables grid cell domination computation.  
We showed that only a small constant number of cells need to be examined, 
which is independent of the dataset cardinality, yielding 
highly efficient skyline computation.  
Our structure also enables parallel computation.  
We thus proposed a parallel skyline algorithm to boost the computation efficiency, 
taking advantage of the parallelization power of GPUs.    
Our cost analysis and experiments confirm the efficiency of the proposed algorithms. 
Our parallel algorithm outperforms state-of-the-art skyline algorithms consistently and by up to over two orders of magnitude in the algorithm response time. 

Our technique also supports parallel processing on CPUs straightforwardly and can be adapted for distributed processing because of its independent grid cell computation. For future work, we plan to design  distributed skyline algorithms on Spark using our structure.
 


\bibliographystyle{IEEEtran}
\bibliography{ms} 

\begin{thebibliography}{10}
\providecommand{\url}[1]{#1}
\csname url@samestyle\endcsname
\providecommand{\newblock}{\relax}
\providecommand{\bibinfo}[2]{#2}
\providecommand{\BIBentrySTDinterwordspacing}{\spaceskip=0pt\relax}
\providecommand{\BIBentryALTinterwordstretchfactor}{4}
\providecommand{\BIBentryALTinterwordspacing}{\spaceskip=\fontdimen2\font plus
\BIBentryALTinterwordstretchfactor\fontdimen3\font minus
  \fontdimen4\font\relax}
\providecommand{\BIBforeignlanguage}[2]{{%
\expandafter\ifx\csname l@#1\endcsname\relax
\typeout{** WARNING: IEEEtran.bst: No hyphenation pattern has been}%
\typeout{** loaded for the language `#1'. Using the pattern for}%
\typeout{** the default language instead.}%
\else
\language=\csname l@#1\endcsname
\fi
#2}}
\providecommand{\BIBdecl}{\relax}
\BIBdecl

\bibitem{bogh2015work}
K.~S. B{\o}gh, S.~Chester, and I.~Assent, ``Work-efficient parallel skyline
  computation for the gpu,'' \emph{PVLDB}, vol.~8, no.~9, pp. 962--973, 2015.

\bibitem{borzsony2001skyline}
S.~Borzsony, D.~Kossmann, and K.~Stocker, ``The skyline operator,'' in
  \emph{ICDE}, 2001, pp. 421--430.

\bibitem{chomicki2003skyline}
J.~Chomicki, P.~Godfrey, J.~Gryz, and D.~Liang, ``Skyline with presorting,'' in
  \emph{ICDE}, 2003, pp. 717--719.

\bibitem{kohler2011efficient}
H.~K{\"o}hler, J.~Yang, and X.~Zhou, ``Efficient parallel skyline processing
  using hyperplane projections,'' in \emph{SIGMOD}, 2011, pp. 85--96.

\bibitem{liu2018skyline}
J.~Liu, J.~Yang, L.~Xiong, J.~Pei, and J.~Luo, ``Skyline diagram: Finding the
  voronoi counterpart for skyline queries,'' in \emph{ICDE}, 2018, pp.
  653--664.

\bibitem{yu2019efficient}
W.~Yu, J.~Liu, J.~Pei, L.~Xiong, X.~Chen, and Z.~Qin, ``Efficient contour
  computation of group-based skyline,'' \emph{IEEE Transactions on Knowledge
  and Data Engineering}, vol.~32, no.~7, pp. 1317--1332, 2020.

\bibitem{zou2008novel}
L.~Zou, L.~Chen, J.~X. Yu, and Y.~Lu, ``A novel spectral coding in a large
  graph database,'' in \emph{Proceedings of the 11th international conference
  on Extending database technology: Advances in database technology}, 2008, pp.
  181--192.

\bibitem{wang2019scalable}
W.~Wang, J.~Zhang, M.-T. Sun, and W.-S. Ku, ``A scalable spatial skyline
  evaluation system utilizing parallel independent region groups,'' \emph{The
  VLDB Journal}, vol.~28, no.~1, pp. 73--98, 2019.

\bibitem{bogh2016skyalign}
K.~S. B{\o}gh, S.~Chester, and I.~Assent, ``Skyalign: A portable,
  work-efficient skyline algorithm for multicore and gpu architectures,''
  \emph{The VLDB Journal}, vol.~25, no.~6, pp. 817--841, 2016.

\bibitem{islam2017computing}
M.~S. Islam, W.~Rahayu, C.~Liu, T.~Anwar, and B.~Stantic, ``Computing influence
  of a product through uncertain reverse skyline,'' in \emph{SSDBM}, 2017, pp.
  1--12.

\bibitem{zois2019complex}
V.~Zois, ``Complex query operators on modern parallel architectures,'' Ph.D.
  dissertation, UC Riverside, 2019.

\bibitem{lougmiri2017new}
Z.~Lougmiri, ``A new progressive method for computing skyline queries,''
  \emph{Journal of Information Technology Research}, vol.~10, no.~3, pp. 1--21,
  2017.

\bibitem{choi2012multi}
W.~Choi, L.~Liu, and B.~Yu, ``Multi-criteria decision making with skyline
  computation,'' in \emph{IEEE 13th International Conference on Information
  Reuse \& Integration}, 2012, pp. 316--323.

\bibitem{bogh2013efficient}
K.~S. B{\o}gh, I.~Assent, and M.~Magnani, ``Efficient gpu-based skyline
  computation,'' in \emph{The 9th International Workshop on Data Management on
  New Hardware}, 2013, pp. 1--6.

\bibitem{zhang2009scalable}
S.~Zhang, N.~Mamoulis, and D.~W. Cheung, ``Scalable skyline computation using
  object-based space partitioning,'' in \emph{SIGMOD}, 2009, pp. 483--494.

\bibitem{nasridinov2017two}
A.~Nasridinov, J.-H. Choi, and Y.-H. Park, ``A two-phase data space
  partitioning for efficient skyline computation,'' \emph{Cluster Computing},
  vol.~20, no.~4, pp. 3617--3628, 2017.

\bibitem{lee2014scalable}
J.~Lee and S.-W. Hwang, ``Scalable skyline computation using a balanced pivot
  selection technique,'' \emph{Information Systems}, vol.~39, pp. 1--21, 2014.

\bibitem{lee2017efficient}
G.~Lee and Y.-H. Lee, ``An efficient method of computing the k-dominant skyline
  efficiently by partition value,'' in \emph{The 3rd International Conference
  on Information Management}, 2017, pp. 416--420.

\bibitem{osm_stat}
\BIBentryALTinterwordspacing
O.~stats report, 2021. [Online]. Available:
  \url{https://www.openstreetmap.org/stats/data_stats.html}
\BIBentrySTDinterwordspacing

\bibitem{kung1975finding}
H.-T. Kung, F.~Luccio, and F.~P. Preparata, ``On finding the maxima of a set of
  vectors,'' \emph{Journal of the ACM}, vol.~22, no.~4, pp. 469--476, 1975.

\bibitem{hose2012survey}
K.~Hose and A.~Vlachou, ``A survey of skyline processing in highly distributed
  environments,'' \emph{The VLDB Journal}, vol.~21, no.~3, pp. 359--384, 2012.

\bibitem{tan2001efficient}
K.-L. Tan, P.-K. Eng, B.~C. Ooi \emph{et~al.}, ``Efficient progressive skyline
  computation,'' in \emph{VLDB}, 2001, pp. 301--310.

\bibitem{kossmann2002shooting}
D.~Kossmann, F.~Ramsak, and S.~Rost, ``Shooting stars in the sky: An online
  algorithm for skyline queries,'' in \emph{VLDB}, 2002, pp. 275--286.

\bibitem{huang2006skyline}
Z.~Huang, C.~S. Jensen, H.~Lu, and B.~C. Ooi, ``Skyline queries against mobile
  lightweight devices in manets,'' in \emph{ICDE}, 2006, pp. 66--66.

\bibitem{kulkarni2019skyline}
R.~D. Kulkarni and B.~F. Momin, ``Skyline computation for big data,'' in
  \emph{Data Science and Big Data Analytics}, 2019, pp. 267--276.

\bibitem{lee2007approaching}
K.~C.~K. Lee, B.~Zheng, H.~Li, and W.-C. Lee, ``Approaching the skyline in z
  order,'' in \emph{VLDB}, 2007, pp. 279--290.

\bibitem{papadias2005progressive}
D.~Papadias, Y.~Tao, G.~Fu, and B.~Seeger, ``Progressive skyline computation in
  database systems,'' \emph{ACM Transactions on Database Systems}, vol.~30,
  no.~1, pp. 41--82, 2005.

\bibitem{sharifzadeh2006spatial}
M.~Sharifzadeh and C.~Shahabi, ``The spatial skyline queries,'' in \emph{VLDB},
  2006, pp. 751--762.

\bibitem{balke2004efficient}
W.-T. Balke, U.~G{\"u}ntzer, and J.~X. Zheng, ``Efficient distributed skylining
  for web information systems,'' in \emph{EDBT}, 2004, pp. 256--273.

\bibitem{zois2018massively}
V.~Zois, D.~Gupta, V.~J. Tsotras, W.~A. Najjar, and J.-F. Roy, ``Massively
  parallel skyline computation for processing-in-memory architectures,'' in
  \emph{The 27th International Conference on Parallel Architectures and
  Compilation Techniques}, 2018, pp. 1--12.

\bibitem{zhu2017parallelization}
H.~Zhu, P.~Zhu, X.~Li, Q.~Liu, and P.~Xun, ``Parallelization of skyline
  probability computation over uncertain preferences,'' \emph{Concurrency and
  Computation: Practice and Experience}, vol.~29, no.~18, p. e4201, 2017.

\bibitem{bogh2017template}
K.~S. B{\o}gh, S.~Chester, D.~{\v{S}}idlauskas, and I.~Assent, ``Template
  skycube algorithms for heterogeneous parallelism on multicore and gpu
  architectures,'' in \emph{SIGMOD}, 2017, pp. 447--462.

\bibitem{mullesgaard2014efficient}
K.~Mullesgaard, J.~L. Pederseny, H.~Lu, and Y.~Zhou, ``Efficient skyline
  computation in mapreduce,'' in \emph{EDBT}, 2014, pp. 37--48.

\bibitem{park2013parallel}
Y.~Park, J.-K. Min, and K.~Shim, ``Parallel computation of skyline and reverse
  skyline queries using mapreduce,'' \emph{PVLDB}, vol.~6, no.~14, pp.
  2002--2013, 2013.

\bibitem{zhang2015efficient}
J.~Zhang, X.~Jiang, W.-S. Ku, and X.~Qin, ``Efficient parallel skyline
  evaluation using mapreduce,'' \emph{IEEE Transactions on Parallel and
  Distributed Systems}, vol.~27, no.~7, pp. 1996--2009, 2015.

\bibitem{chester2015scalable}
S.~Chester, D.~{\v{S}}idlauskas, I.~Assent, and K.~S. B{\o}gh, ``Scalable
  parallelization of skyline computation for multi-core processors,'' in
  \emph{ICDE}, 2015, pp. 1083--1094.

\bibitem{github}
\BIBentryALTinterwordspacing
GitHub, 2021. [Online]. Available: \url{https://github.com/chiewen/SkyCell.git}
\BIBentrySTDinterwordspacing

\bibitem{osm}
\BIBentryALTinterwordspacing
OpenStreetMap, 2021. [Online]. Available: \url{https://www.openstreetmap.org/}
\BIBentrySTDinterwordspacing

\end{thebibliography}

\end{document}